\documentclass[12pt,preprint2,useAMS,usenatbib,trackchanges]{aastex61}

\usepackage{rotating,times,pictex,graphicx,latexsym}
\usepackage{color}
\usepackage{longtable}
\usepackage{amsmath}
\usepackage{nth}
\usepackage{array}

\def\apjl{ApJ}%
\def\apjs{ApJS}%
\def\ao{Appl.~Opt.}%
\def\aap{A\&A}%
\newcommand{\mic}{$\mu$m}
\newcommand{\kms}{km\,s$^{-1}$}
\newcommand{\dg}{$^{\circ}$}

\newcommand{\Lsolar}{L$_{\odot}$}
\newcommand{\htwo}{H$_2$}
\newcommand{\ha}{H$\alpha$}
\newcommand{\feii}{$\left[\right.$Fe\,{\sc ii}$\left.\right]$}
\newcommand{\hii}{H\,{\sc ii}}
\newcommand{\sii}{$\left[\right.$S\,{\sc ii}$\left.\right]$}

\newcommand{\Wm}{$\times$\,$10^{-18}$\,W\,m$^{-2}$}

\shorttitle{YSO jets in the Galactic Plane}
\shortauthors{Makin and Froebrich}

\submitjournal{the Astrophysical Journal Supplement Series}
\accepted{July 2017}

\begin{document}

\title{YSO jets in the Galactic Plane from UWISH2:\\ IV - Jets and Outflows in Cygnus-X}

\author{S.V.\,Makin}
\affiliation{Centre for Astrophysics and Planetary Science, University of Kent, Canterbury, CT2 7NH, United Kingdom}
\author{D.\,Froebrich}
\affiliation{Centre for Astrophysics and Planetary Science, University of Kent, Canterbury, CT2 7NH, United Kingdom}


\email{svm67@kent.ac.uk}

\begin{abstract} 
We have performed an unbiased search for outflows from young stars in Cygnus-X using 42\,deg$^2$ of data from the UKIRT Widefield Infrared Survey for \htwo\ (UWISH2 survey), to identify shock-excited near-IR \htwo\ emission in the 1\,--\,0\,S(1) 2.122\,\mic\ line. We uncovered 572 outflows, of which 465 are new discoveries, increasing the number of known objects by more than 430\,\%. This large and unbiased sample allows us to statistically determine the typical properties of outflows from young stars.

We found 261 bipolar outflows and 16\,\% of these are parsec-scale. The typical bipolar outflow is 0.45\,pc in length and has gaps of 0.025 to 0.1\,pc between large knots. The median luminosity in the 1\,--\,0\,S(1) line is 10$^{-3}$\,\Lsolar. The bipolar flows are typically asymmetrical, with the two lobes misaligned by 5\dg, one lobe 30\,\% shorter than the other, and one lobe twice as bright as the other. Of the remaining outflows, 152 are single-sided and 159 are groups of extended, shock-excited \htwo\ emission without identifiable driving sources. Half of all driving sources have sufficient WISE data to determine their evolutionary status as either protostars (80\,\%) or classical T-Tauri stars (20\,\%). One fifth of the driving sources are variable by more than 0.5\,mag in the K-band continuum over several years. 

Several of the newly-identified outflows provide excellent targets for follow up studies. We particularly encourage the study of the outflows and young stars identified in a bright-rimmed cloud near IRAS\,20294$+$4255, which seems to represent a textbook example of triggered star formation.
\end{abstract}

\keywords{catalogs; ISM: jets and outflows stars: formation; stars: protostars; stars: winds, outflows; surveys} 

\section{Introduction} \label{sec:intro} 
Jets and outflows can be considered a natural by-product of the star formation process, with the infall of material from accretion discs around Young Stellar Objects (YSOs) simultaneously powering the ejection of gas and dust in collimated streams along the axis of the star's rotation. These outflows interact with the local environment via collisions generating fast shocks (10\,--\,100\,\kms) which can be detected at much shorter wavelengths than the forming stars, making outflows incredibly useful signposts of early-stage and ongoing star formation. In the almost 40 years since they were first correctly recognised \citep{snell1980} much work has been done to understand the interplay between the accreted and ejected mass, the jet launching, collimation and propagation, the energetics and timescales involved, and the role of feedback from outflows in the star formation process (see the reviews of e.g., \citet{bachiller1996}, \citet{bally2007}, \citet{frank2014}).

Despite the enormous progress that has been made thus far, it is still not clear which factors govern the properties of the outflows, such as the length, luminosity, and orientation. Do emission features (knots) form as a result of multiple accretion events ejecting material at different velocities, or by interactions with the local environment - or a mixture of both? Is the mass of the star, the evolutionary stage, or the proximity of neighbours to either trigger or disrupt the star formation process (via momentum feedback and turbulent energy) the most important among the stellar properties? How much of a role does the environment in the natal cloud play on large scales? In order to investigate these questions, we require large, unbiased statistical samples of young stellar outflows from across the Galactic Plane. 

To establish just such a sample, the UKIRT Widefield Infrared Survey for \htwo\ (UWISH2) survey was carried out in order to image the Galactic Plane (\citet{froebrich2011}, hereafter F11), with two extensions in 2013 to cover the Auriga and Cassiopeia, and Cygnus-X regions (\citealt{froebrich2015}, hereafter F15). The survey used a narrow-band filter centred on the molecular hydrogen ro-vibrational transition 1\,--\,0\,S(1) at 2.122\,\micron, which is produced via the shock excitation of hot, dense gas (T\,$\sim$\,2000\,K, n\,$\geq$\,$10^3$\,cm$^{-3}$), and is an excellent tracer of shocks from YSO outflows, supernovae, and planetary nebulae. \htwo\ has been used to trace outflows since  \citet{garden1990}, but the first unbiased survey was a pioneering work by \citet{stanke2002} in Orion\,A. Since then, hundreds of outflows have been detected across many regions in the Galactic plane, e.g. in \citet{jiang2004}, \citet{walawender2005}, \citet{hatchell2007}, \citet{davis2007,davis2008,davis2009}, \citet{varricatt2010}, \citet{khanzadyan2012}, \citet{walawender2013}, and \citet{bally2014}, whilst \citet{lee2012,lee2013} have investigated \htwo\ outflows from Spitzer-detected Extended Green Objects.

Near-infrared (NIR) tracers allow us to mitigate the effects of extinction that can obscure outflows in optical lines such as \ha\ and \sii\, but with greatly improved spatial resolution over other molecular transitions such as CO and SiO. Other NIR transition lines have also been used in surveys to find shocks from outflows, such as the 1.644\,\micron\ line of \feii\ used in the UWIFE survey \citep{lee2014}, the sister survey to UWISH2. Using combinations of lines can probe the interplay between different shock velocities and the environment the outflows are interacting with. 

Using the UWISH2 data, the outflows in Serpens and Aquila (\citet{ioannidis2012a,ioannidis2012b}, hereafter IF12a and IF12b respectively), and Auriga and Cassiopeia (\citet{froebrich2016}, hereafter F16) have been catalogued so far, with the remainder of the Galactic Plane yet to be analysed. In this paper we analyse the outflows in the Cygnus-X region, which represents a rich, high-mass star-forming complex, home to a number of \hii\ regions and young OB associations. The regions investigated in F16 represent low-mass star-forming regions in the outer Galaxy, whilst the regions investigated in IF12a and IF12b slice through the inner Galactic Plane between 18\dg\,$<l<$\,35\dg\, and thus should represent the typical star formation environment in the Galactic Plane. In our follow-up paper (Makin \& Froebrich, 2017 \textit{in prep.}) we will perform the first statistical analysis to compare these regions against each other and probe the effects of these differing environments on the star formation process with a large and unbiased sample. 

This paper catalogues the outflows from YSOs in the Cygnus-X region as Molecular Hydrogen emission-line Objects (MHOs; \citet{davis2010}) based on the data from the UWISH2 survey. In Section \ref{sec:analysis} of this catalogue, we detail our methods and the UWISH2 data, whilst in Section \ref{sec:results} we present and discuss the results, in terms of the outflow and driving source properties, and also include a list of the newly identified candidate star clusters that were discovered as a result of this work. 

\section{Data analysis} \label{sec:analysis}

\begin{figure}
  \center
    \includegraphics[width=\columnwidth]{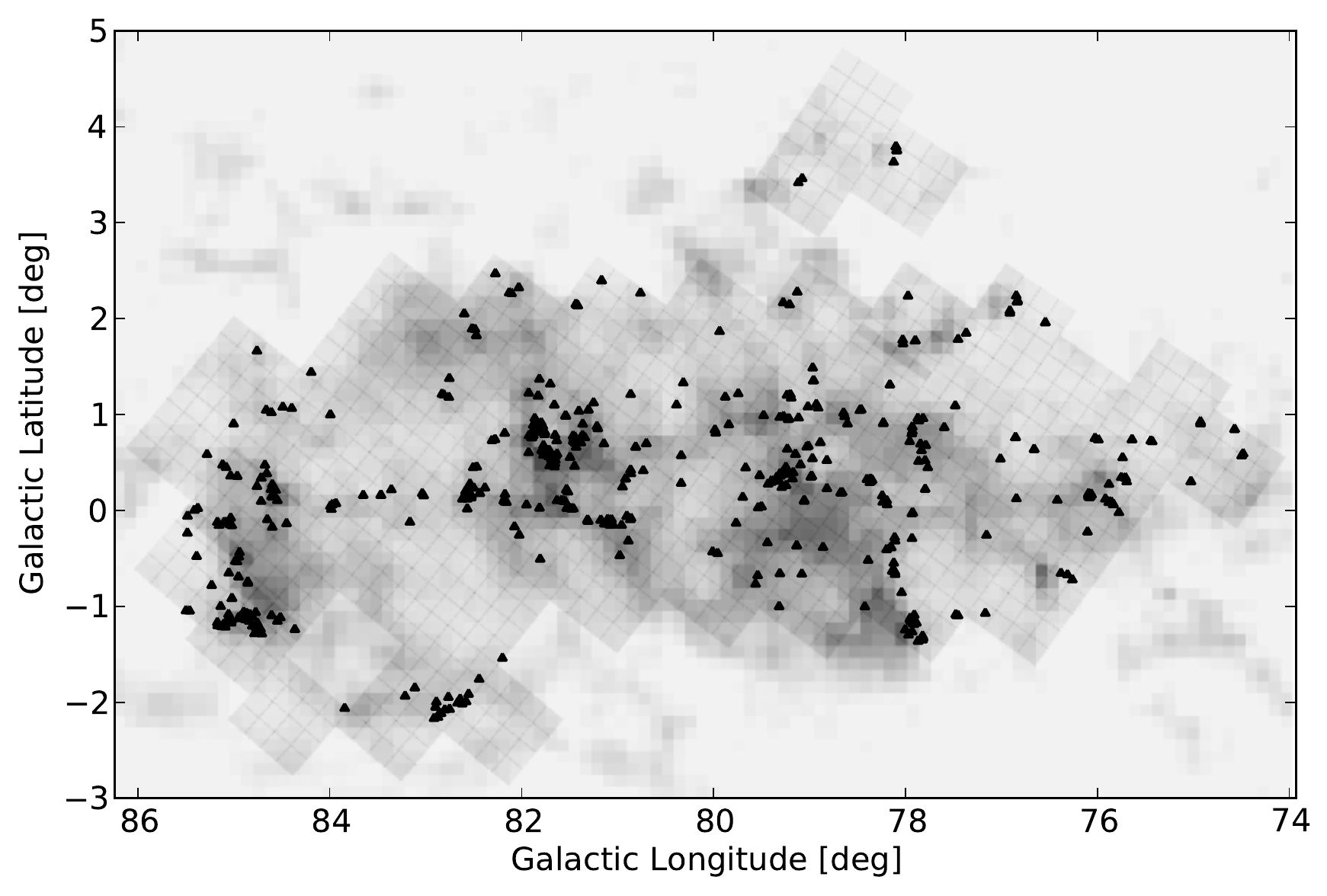} 
    \caption{\label{fig:coverage_map} A CO map of the Cygnus-X region of the Galactic Plane from \citet{dame2001}, where higher levels of CO emission are represented with darker shades. We overlay a grid to demonstrate the coverage of the UWISH2 survey tiles in this region, representing an area of 41.99 square degrees. The black triangles mark the positions, in Galactic Coordinates, of all the outflows we detected.}
\end{figure} 

\subsection{The UWISH2 data}
We utilised data from UWISH2, specifically the extension toward the Cygnus-X region of the Galactic Plane as described in F15. Cygnus-X was observed between April \nth{6} and December \nth{11} in 2013. The survey used the Wide Field Camera (WFCAM; \citet{casali2007}) to obtain images in the 1\,--\,0\,S(1) narrow-band filter at 2.122\,\micron\ ($\Delta \lambda$\,=\,0.021\,\micron), at the UK Infra-Red Telescope (UKIRT) in Hawaii. A total exposure time of 720\,s per pixel was used, and micro-stepping during the observations gave a final pixel size of 0.2\arcsec\,$\times$\,0.2\arcsec. The typical 5$\sigma$ surface-brightness detection limit is 4.1\,$\times$\,10$^{-19}$\,W\,m$^{-2}$\,arcsec$^{-2}$ when averaged over the typical seeing in the data, which is 0.8\arcsec\ (F15). We also used J, H and K-band images obtained from the UKIRT Infrared Deep Sky Survey (UKIDSS; \citet{lawrence2007}) Galactic Plane Survey (UGPS; \citet{lucas2008}). UGPS used the same instrument and tiling set-up as the UWISH2 survey, hence the data from both surveys are consistent in terms of their quality and resolution. The UGPS per-pixel exposure time in the K-band, used for continuum subtraction, is 40\,s.

In the extension toward Cygnus-X, the UWISH2 survey covered the approximate region from Galactic longitudes 74\dg\,$<l<$\,86\dg\ and Galactic latitudes -3\dg\,$<b<$\,5\dg. The layout of the UWISH2 coverage in this region is demonstrated in Fig.\,\ref{fig:coverage_map}, and in total, represents an area of 41.99 square degrees. In order to isolate the narrow-band \htwo\ emission we performed continuum-subtraction using K-band continuum images (K-filter at 2.201\,\mic\ with a 0.34\,\mic\ bandwidth) and applying psf-fitting to correct the seeing, following \citet{lee2014}.  In addition to \htwo-K images, we also constructed RGB images using \htwo-, K- and J-band data. All images (J, H, K, \htwo, and the \htwo-K difference images) are publicly available in FITS format on the UWISH2 website\footnote{The UWISH2 Survey: \url{http://astro.kent.ac.uk/uwish2/index.html}}.

\subsubsection{\htwo-K and JK\htwo\ images} 
We examined the \htwo-K and JK\htwo\ images simultaneously in order to visually identify shock-excited emission-line features (knots) that likely belong to jets and outflows, based on the morphology and colour of the knots. Radiatively-excited or fluorescing cloud edges, as well as image artefacts, are more readily identified when using both images in conjunction. Shock-excited \htwo\ emission is bright (positive-valued) in the \htwo-K difference images, and appears red in the colour JK\htwo\ images. Whilst fluorescently-excited cloud edges can be mistaken for shock-excited emission in the \htwo-K images, they appear brownish in the JK\htwo\ RGB images due to the inclusion of additional excited emission-lines in the J- and K-band continuum filters. This makes it easy to distinguish between fluorescently-excited cloud edges and shock-excited emission when both images are used for the identification. 

Artefacts from very bright stars, such as electronic cross-talk, form donut-shaped rings or arc-shaped features that can look like bow-shocks in the \htwo-K images, but tend to be comprised of a combination of superposed coloured rings in the colour images. Some artefacts may appear as bright (positive-valued) extended objects in the \htwo-K difference images, but appear dark in the colour images, instead of being red. All of these artefacts, and the cloud edges, are rejected as not being real shock emission, and we exclude them from both this catalogue and our analysis. Examples of these can be seen in Fig.\,5 of F15. 

We found that frequently, the outflows can be situated on image edges or corners, and may span two or more images along their length. In order to be able to view these outflows in full, we used the Montage\footnote{Montage: \url{http://montage.ipac.caltech.edu/}} software to create composite images for each of these cases. 

\begin{figure*}
	\center
    	\includegraphics[width=0.77\linewidth]{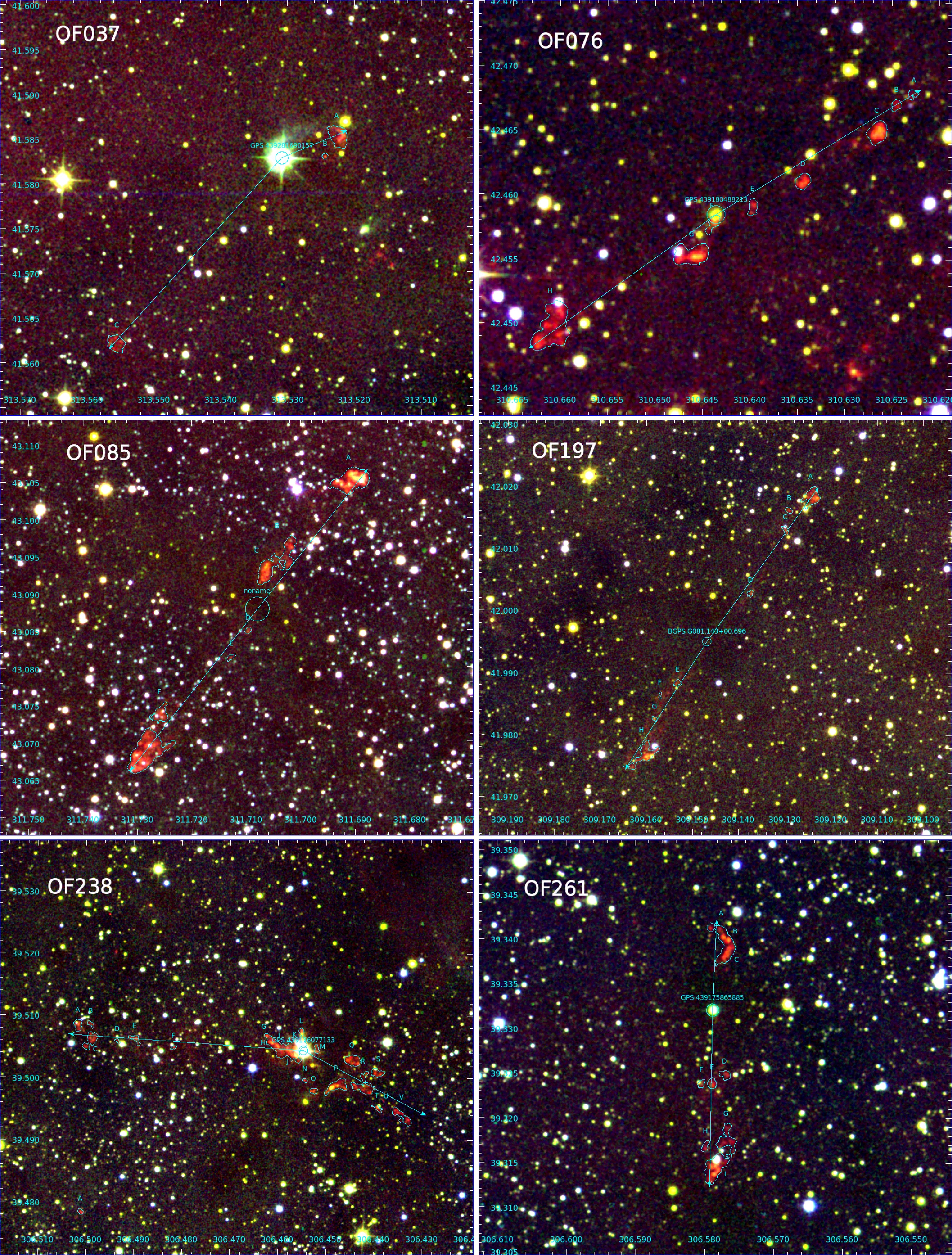}
        \caption{\label{fig:f2} A selection of previously unknown, \htwo-detected outflows discovered as a result of this survey. Outflow lengths are calculated using a distance of 1.4\,kpc towards all objects. Full-size, high-resolution versions of each image are available both in the online journal and in the MHO catalogue.
        \textbf{Top left:} OF\,037 (MHO\,4044), a 0.9\,pc outflow, likely from variable star V1219\,Cyg; 
        \textbf{Top right:} OF\,076 (MHO\,3972), a 0.9\,pc outflow known from CO observations ([GKM2012b]\,G82.186+0.105); 
        \textbf{Centre left:} OF\,085 (MHO\,4005), a 1.2\,pc outflow perpendicular to a dark filament with an embedded driving source; 
        \textbf{Centre right:} OF\,197 (MHO\,3878), a 1.3\,pc outflow known from CO observations ([GKM2012b]\,G81.140+0.687) with an embedded driving source; 
        \textbf{Bottom left:} OF\,238 (MHO\,3493), a 1.5\,pc complex outflow, more-or-less perpendicular to a dark filament. The bright star identified as potential driving source is the emission-line star [D75b]\,Em*\,20-090, also known as [MSX6C]\,G077.9280+00.8711;
        \textbf{Bottom right:} OF\,261 (MHO\,3497), a bright 0.7\,pc outflow that appears to run along a dark filament.}
\end{figure*}

\subsection{Identifying outflows and their driving sources}
\subsubsection{The UWISH2 catalogue data}
As our basis, we used the subset of objects classified as ``jet'' emission-line features falling within the Cygnus-X survey area, as listed in the catalogue of extended \htwo\ emission-line sources from UWISH2 (F15). The UWISH2 catalogue groups individual objects of the same classification if they are closer together than 0.1\dg\ (corresponding to 2.4\,pc at an assumed distance of 1.4\,kpc). There are 210 such groups of ``jet'' emission in Cygnus-X, comprising 30\,\% of the total number of jet groups in the full UWISH2 catalogue. Emission is classified as ``jet'' in F15 if the morphology suggests shock-excited emission which is either recognisably an outflow, or likely part of an outflow (including isolated bow shocks and individual knots). 

\subsubsection{Assigning emission-line objects to outflows} 
Within each Cygnus-X group of jet features, we divided the emission-line features (``knots'') into likely outflows based on factors such as the knot morphology and alignment, and gave each one an outflow ID number in the format [OF\,XXX]. For every group, we attempted to minimise the number of outflows we defined, particularly in complex regions without clear solutions. This means that where there were contiguous collections of knots, or where knots formed a chain but had no clear driving source, we assigned one outflow ID number to that collection of knots, rather than giving each knot its own ID number. We also included knots which did not have a high enough surface brightness to warrant inclusion in the UWISH2 catalogue. This typically applied to one or more knots per outflow, in line with what was found in Auriga and Cassiopeia (F16).

When grouping the knots into outflows, we followed the methods described in the catalogue of Molecular Hydrogen emission-line Objects (MHOs; \citealt{davis2010}), so that each newly-discovered outflow is also given a new MHO number in the format [MHO\,XXXX] and included in the MHO catalogue\footnote{The MHO Catalogue, \url{http://astro.kent.ac.uk/~df/MHCat/index.html}}. Since Cygnus-X is a well-studied region of the Galactic Plane, there are 147 already-known MHOs falling within our survey area. Therefore, if the outflow we identify has already been discovered and has an entry in the MHO catalogue, we keep the MHO ID number that has already been assigned, and simply add any additional knots we have detected to it. In some regions, the association of \htwo\ features to outflows is not straightforward and hence there are some outflows comprised of more than one already-known MHO, particularly where each lobe in an outflow has previously been assigned separate MHO numbers. In the discussion section of this paper we will, for clarity, refer to both our assigned outflow ID number, and any corresponding MHO numbers when referring to any given outflow. Our data table (Appendix Table\,\ref{sec:appx_mhodata}) and set of images (Appendix\,\ref{sec:appx_images}) also follow this convention and list all corresponding MHO ID number(s) for each outflow.  

\subsubsection{Identifying driving sources}  
We utilised a selection of all-sky point-source catalogues in order to identify potential driving sources for our outflows (see Sect.\,\ref{sec:ps_cat} for further details). Using all-sky catalogues enables us to be consistent with analysis of outflows in Auriga and Cassiopeia (F16) and Serpens and Aquila (IF12a). Source candidate selection was based largely on alignment with the outflow axis, with preference given to those sources which had longer-wavelength catalogue entries, and to those with reflection nebulae extended toward the outflow lobes. Where multiple driving source candidates aligned with the outflow axis, we considered the degree of reddening and variability of each candidate in the near- and mid-infrared in order to inform our decision. We assigned each outflow a percentage ``confidence'' level in the selected source, such that those outflows with multiple possible sources are given a low confidence, and isolated outflows with one good candidate source are given a high confidence. 

There were some instances where a driving source could not be properly identified, i.e.: 
\begin{itemize}
\item Outflows comprised of a clear chain of knots and an obvious driving source location, but no catalogue detections. For these, we recorded the location coordinates and called the source ``\textit{noname}'', as was done in Auriga and Cassiopeia (F16), and Serpens and Aquila (IF12a). 
\item Outflows comprised of a clear chain of knots, but with neither catalogue detections nor obvious driving source locations. For these, we recorded coordinates at a reasonable location along the axis and called the source ``\textit{unknown}''. We gave these sources a low confidence level. 
\item Groups of individual, uncollimated knots. We called the source ``\textit{unknown}'' and gave them a confidence level of zero to exclude them from all analysis. We record their coordinate location as the mean position of all the knots.      
\end{itemize}

\subsection{Measuring the outflow properties}
We noted whether each outflow is bipolar (both lobes are detected), single-sided (only one lobe is detected), or a single knot (this includes pairs and collections of knots with no clear source). The properties of each lobe were measured separately.

We measured the length of each outflow lobe from the coordinates of the assigned driving source, out to the furthest tip of the most distant knot in that lobe. We then measured the position angle from the north direction towards east. Each outflow sub-feature was labelled sequentially, starting with `A' at the knot furthest from the source. For bipolar outflows, this began from the end of the lobe with the smallest position angle, toward the lobe with the largest position angle. Sub-features in groups of individual knots were not labelled in any particular order.  

In each outflow lobe we measured the gaps and position angles between sequential knots (or clusters of knots in very complex bow shocks). This only applies to single-sided or bipolar outflows; groups of individual emission knots were excluded. We drew manual contours around each emission-line knot using the \htwo-K images. This ensured that all possible \htwo\ emission in an outflow was captured (including where the knots were too faint to be included in the UWISH2 catalogue), while excluding emission contributions from bright stars and radiative emission from cloud edges. We also measured the local background flux near each outflow by drawing a contour around a nearby region containing no \htwo\ emission. We then measured the integrated flux inside each contour, and calibrated the fluxes in the same way as described in F15. In Appendix Table\,\ref{sec:appx_mhodata} we show the total flux for each outflow lobe (in the case of collections of single knots, we show the total flux from all the knots). Note that no correction was made for additional line fluxes from other lines in the K-band filter used for the continuum subtraction since the precise corrections required depend on the excitation conditions (such as temperature, shock speed, density and non-shock excitation mechanisms). Therefore there is a systematic underestimation in the fluxes of about 30\,\% on average. However, this will not change our results as we investigate only the statistical distributions, and assume that all shocks are produced in similar conditions.

\subsection{Measuring the properties of the driving sources}
\subsubsection{The point-source catalogues} \label{sec:ps_cat}
We checked every candidate driving source against: UGPS \citep{lucas2008}; 
the Two Micron All Sky Survey (2MASS, \citealt{skrutskie2006}); 
the AllWISE data release (\citealt{cutri2013}); 
the AKARI\,/\,IRC mid-infrared all-sky survey (\citealt{ishihara2010}); 
the AKARI\,/\,FIS All-Sky Survey Point Source Catalogues (\citealt{yamamura2010}); 
the Bolocam Galactic Plane Survey (BGPS, \citet{bgps2010}); 
and Infrared Astronomical Satellite (IRAS, \citet{iras1984}) point-source catalogues. 

Where a driving source was detected in one (or more) of these catalogues, we logged the ID number(s). For UGPS, 2MASS and WISE, we also recorded the magnitudes in each filter associated with the ID(s). For AKARI, BGPS and IRAS detections there can be a large offset in the coordinates, so we avoided attributing these IDs to a source if the detection was too distant from the outflow. Where there were multiple YSOs in a small group, the flux from these far-infrared and sub-mm detections is most likely attributable to the whole group and hence we did not record the fluxes from these catalogues. In Appendix Table\,\ref{sec:appx_mhodata} we show where an outflow has a detection in each of these catalogues, and the coordinates listed are those associated with the selected driving source. Where the source is detected in UGPS we preferentially use that ID and coordinate position. If the source has no detection in UGPS, we use the ID and coordinates of the shortest-wavelength catalogue in which it is detected. Where there is no identifiable driving source, we use the mean coordinate position of the \htwo\ emission-line knots that comprise the outflow.

\subsubsection{Near- and mid-infrared magnitudes} \label{sec:MIR}
We recorded the NIR magnitudes from UGPS and 2MASS (shown in Appendix Table\,\ref{sec:appx_mhodata}) for all outflow driving sources that have detections in those catalogues. We calculated the J-H and H-K colours for all sources with magnitudes in at least one of these catalogues, and the variability for all sources that have at least two K-band magnitudes.

From WISE, we logged the mid-infrared (MIR) magnitudes (shown in Appendix Table\,\ref{sec:appx_mhodata}). From this we calculated the slope, $\alpha$, of the spectral energy distribution (SED) for each source with an AllWISE detection. We used all four WISE bands between 3.4\,\micron\ and 22\,\micron, following the method detailed in \citet{majaess2013}. In our analysis, we compare the effects of the evolutionary stage on various outflow properties; the youngest/least evolved objects are protostars ($\alpha > 0$) and the oldest/more evolved objects are classical T-Tauri stars (CTTSs, $\alpha < 0$). 

\subsubsection{Environment} \label{sec:environment}
To investigate the environmental dependence of outflow properties, we noted the type of environment that each outflow appears to originate from. It must be stated that in this paper, we use the word ``cluster'' rather loosely and not solely in the traditional meaning of Open and Galactic clusters. Where we refer in Appendix Table\,\ref{sec:appx_mhodata} to ``clustered vs. non-clustered'' environments, we intend that ``clustered environment'' describes any area of ongoing star-formation activity, including star forming regions, \hii\ regions, compact groups of reddened stars and known YSOs, particularly those with large amounts of circumstellar and reflection nebulosity. Non-clustered refers to any environment where the outflows and YSOs are more widely spread and generally isolated. 

In Appendix Table\,\ref{sec:appx_clusters} we list all the previously unknown candidate clusters and groups of YSOs identified in the near-infrared JHK images that are associated with, or are in proximity to, our outflows. We did not perform a systematic search for clusters in our survey and hence there are likely more yet to be discovered. In this table we refer to ``clusters'' and ``groups'' to describe the general appearance of the collections of stars. In this usage, ``cluster'' (as opposed to ``clustered'' as described above) refers to an unknown and reasonably circular collection of stars, irrespective of the number of apparent members, and a ``group'' refers to a collection of stars that show a more loose and filamentary structure. We also estimate the number of members visible in our JHK images, any outflow IDs associated with it, and further comments on each, such as whether or not the cluster is associated with any known IRAS objects, \hii\ regions or star forming regions.

\subsubsection{Distances to outflows}
In Cygnus-X we look down a spiral arm \citep{wendker1991}, with complex interlayering of molecular clouds superimposed onto each other along the line of sight. The typical range of distances in Cygnus-X can be anywhere from $\sim$\,700\,pc toward the Cygnus Rift, up to a few kpc toward e.g. Westerlund\,1 ($\sim$\,4\,kpc, \citet{kothes2007}), but distances can be much further. This means that attempting to ascertain accurate distances to our outflows is beyond the scope of this paper. Instead, we use the commonly adopted distance of 1.4\,kpc toward all the outflows in our survey (based on e.g., the Red MSX survey \citep{lumsden2013}, the maser parallax study of \citet{rygl2012}, and other studies such as \citet{varricatt2010}, \citet{rivera2015}, \citet{kmh2014}). This does mean that the error margins on the distances toward our outflows may be rather large, and if the outflows are systematically located further away than this then the lengths and luminosities we report will be underestimated. We will investigate the extent to which our distributions are affected by this in a follow-up paper with better-constrained distances toward individual outflows, where these can be determined.

\subsubsection{Statistical testing} \label{sec:kstest}
In our analysis, we investigated the bulk properties of our outflows by dividing each of our samples (e.g., the lobe lengths) into sub-samples according to some other parameter (e.g., whether the driving source is a protostar or CTTS, or whether it originates from a clustered or isolated environment). We then performed 2-sample Kolmogorov-–Smirnov (KS) tests of these sub-samples against each other, with a null hypothesis that they are drawn from the same parent distribution. We accept the null hypothesis if the probability $p >$\,90\,\% (i.e., they are drawn from the same distribution) and reject it if $p <$\,10\,\% (they are drawn from different distributions). Where we refer to the KS test in the results and discussion section, we provide the percentage probability value. If the test results in a probability 10\,\%\,$< p <$\,90\,\%, the test is considered inconclusive. 

\section{Results and Discussion} \label{sec:results}

\subsection{The Cygnus-X outflow catalogue} 
The spatial distribution of the outflows we defined in Cygnus-X is demonstrated in Fig.\,\ref{fig:coverage_map}. We mark the coordinate position of each outflow with a black triangle on top of the CO map from \citet{dame2001}, where we have also overlaid a grid showing the coverage of the UWISH2 survey in the region. Clearly, the majority of our outflows are situated within high column density CO features.
 
In total we identified 572 outflows, 261 of which (46\,\%) are bipolar outflows, 152 (27\,\%) are single-sided outflows (i.e. with only one lobe visible), and 159 (28\,\%) are individual knots or collections of knots. Of our 572 outflows, 107 (19\,\%) are comprised (in part or wholly) of objects already known in the MHO catalogue. Of the already-known MHOs that fall inside our survey area, all but one are accounted for in our survey. MHO\,3411 is the only exception: although it was detected, it appears in our images as a point source rather than an extended \htwo\ emission-line feature, and hence was excluded from our list. This catalogue therefore represents an addition of 465 entirely new outflows, which is the single largest collection of newly identified \htwo\ outflows published to date (and an increase of 435\,\% over the 107 previously known outflows in our survey area).

Out of the 210 groups of ``jet'' emission features originally defined in F15, ten were found to contain only fluorescently-excited cloud edges and hence have been rejected. Perhaps unsurprisingly, the group containing the largest number of outflows was the one comprising the DR\,21 region, with 38 outflows (although the majority of these are already known in the MHO catalogue from \citet{davis2010}). The median number of outflows per group is 1, and on average we find 2.8 outflows per group in Cygnus-X and 2 outflows per group in Auriga and Cassiopeia (F16). There are 450 groups across the Galactic Plane which are yet to be investigated: we estimate that the number of undiscovered outflows in UWISH2 may be anywhere between 900 and 1200. Since the UWISH2 survey has yielded a total of 801 outflows so far from the regions that have been investigated, we estimate that the final total of outflows from UWISH2 will be up to 2000. 

In Appendix Table \ref{sec:appx_mhodata} we present the outflows we defined, along with the following properties: (1) the Outflow number we assigned; (2) the MHO number, as listed in the MHO catalogue; (3-4) the Right Ascension and Declination in J2000 of the outflow driving source; (5-6) the length of each lobe measured in degrees; (7-8) the position angle of each lobe measured in degrees; (9-10) the total flux in the lobe(s); (11) whether the outflow seems to originate from a cluster (Y) or not (N); (12) whether the outflow is bipolar (B), single-sided (S) or a single knot/collection of knots (K); (13) the identifier of the most likely driving source; (14) the confidence level (\%) that the source we have identified is correct; (15-18) NIR magnitudes from UGPS; (19-21) NIR magnitudes from 2MASS; (22-25) MIR magnitudes from AllWISE; and (26) the point-source catalogues in which the driving source is detected (G for UGPS, 2 for 2MASS, W for AllWISE, A for AKARI, B for Bolocam GPS). 

The corresponding images for each outflow are found in a figure set in the online journal, an example of which can be found in Appendix \ref{sec:appx_images}. Some of the outflows are also detected at other wavelengths. Of our outflows, 27 (4.7\%) are associated with known Herbig-Haro (HH) objects, and 38 (6.6\%) have been measured as CO outflows in \citet{gkm2012}. Where we find that our outflows have been detected in other surveys, we list the relevant ID numbers in the caption for the image of that outflow in the figure set.

\subsection{Driving source properties}

\begin{figure}
	\center
    	\includegraphics[width=\columnwidth]{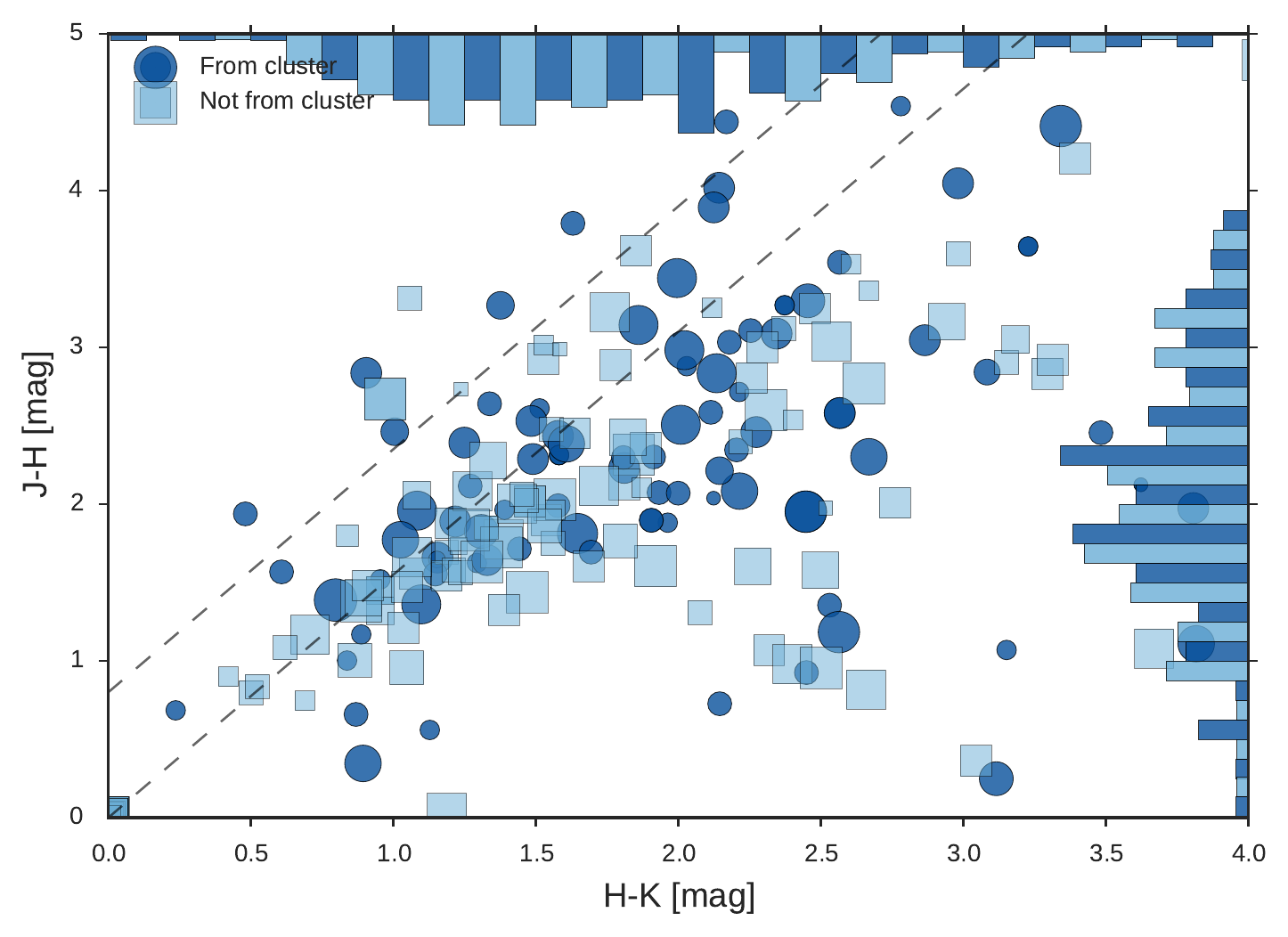} 
    	\caption{A near-infrared colour-colour diagram using data from both UGPS and 2MASS. Where a detected driving source has data missing from one or more of the UGPS bands, or the source is saturated in UGPS we use the 2MASS magnitudes instead (where available). The size of the scatter points increases with our confidence in the assigned driving source. The dark circles represent sources located in clustered environments, whilst the pale squares signify sources that are more isolated. A number of objects sit above the dashed lines representing the reddening band for normal stellar photospheres, and these are discussed in more detail in Sect.\,\ref{sec:nir}. 
        \label{fig:NIR_CCD}}
\end{figure}

For the bipolar outflows, we have identified potential driving sources for 215 (82\%) of them, and likely driving source locations for 27 (10\,\%). By ``driving source location'' we refer to the cases where there are no detections in our selected catalogues, but we can still make a reasonable estimation of the most likely location for the driving source. For instance, where two bow shocks are travelling away from each other so we can suggest that the source lies between them, or where there is a dark region of cloud bisecting the outflow axis suggesting a potentially embedded driving source. For the single-sided outflows, 146 (96\,\%) have been assigned driving sources. In some cases, more than one outflow may appear to originate from the same driving source e.g., OF\,327 (MHO\,3955) and OF\,328 (MHO\,3956). These ``multi-polar'' outflows will be discussed in more detail in Sect.\,\ref{sec:orientation}.

\subsubsection{Clustered vs. non-clustered environments}
For the small fraction of outflows where we couldn't identify a reasonable driving source, it is still possible to judge whether or not the outflow likely originates from somewhere in a cluster of stars, or is more isolated. Since most stars form in clusters, and the majority of protostars ($\alpha > 0$) and CTTSs ($\alpha < 0$) are located in clusters, it is reasonable to expect that most of the outflows driven by young stars would be found in or around clusters. What we actually found in Cygnus-X is that 238 (42\,\%) of \htwo\ outflows are associated with clusters, compared to 334 (58\,\%) which are isolated. The same result was found in Auriga and Cassiopeia (F16; 41\,\% clustered vs. 59\,\% non-clustered). It may be that this result represents an evolutionary trend, i.e. that sources visible in NIR clusters are older than the more isolated sources. We see a higher proportion of young driving sources outside of clusters; in Cygnus-X 34\,\% of our young driving sources are found in clusters, but 46\,\% are isolated (compared to 28\,\% and 41\,\% respectively, in Auriga and Cassiopeia). Since these fractions are so similar, it is unlikely to be a selection effect (i.e., that we cannot see the clusters in Auriga and Cassiopeia simply because they are further away).

\subsubsection{Near- and mid-infrared magnitudes} \label{sec:nir}
Of the 413 outflows that are either bipolar or single-sided, 269 (65\,\%) have candidate driving sources with detections in all four WISE filters. When we calculate the slope of the SED ($\alpha$), 215 (80\,\%) of the WISE-detected sources have a positively sloping SED, and 54 (20\,\%) have a negatively sloping SED, suggesting that the majority of our driving sources are protostars. 

We checked the NIR colours of the bipolar and single-sided driving sources, 294 of which (71\,\%) have detections in either UGPS, 2MASS, or both. Figure\,\ref{fig:NIR_CCD} shows the NIR colour-colour diagram of these driving source candidates, where we preferentially use UGPS data if there are magnitudes in J-, H- and K-bands (due to the improved resolution), else we use 2MASS. We also use 2MASS data if the K-band magnitudes are brighter than 10\,mag in UGPS, which accounts for the saturation limit. Although most of the sources are along or under the reddening band, there is a considerable amount of scatter. We investigated the possible reasons for this and find that generally, it either occurs with sources in star-forming regions (such as in the vicinity of DR\,21 or other dusty areas) or is caused by variability. There are also some cases of potential binary systems (such as with OF\,448 and OF\,449, already-known MHO\,3400), and some cases where the 2MASS data has been used because UGPS is missing data from one or more filters, but the UGPS images show two or more very close stars that are unresolved in 2MASS. 

\subsubsection{Near-infrared variability}
Since we had up to three K-band magnitudes for 294 of the sources, we investigated the NIR variability within the K-band. Between the K$_1$ and K$_2$ bands in UGPS, we find that 267 objects have detections in both epochs; 62\,\% (166 objects) are variable by more than 0.1\,mag, and 19\,\% (50 objects) by more than 0.5\,mag. Only 9\,\% of candidate sources are variable by more than 1.0\,mag (24 objects). There is no statistical difference between objects that are increasing or decreasing in magnitude between the two epochs of UGPS. 

By contrast, there are 168 source candidates with detections in both 2MASS (K$_{\sc S}$) and the first epoch (K$_1$) in UGPS. Over the typical timescale of several years, 121 of these 168 (72\,\%) are variable by more than 0.1\,mag, and 52 (31\,\%) have a greater variability than 0.5\,mag. Thirty-four of the objects show more than 1.0\,mag of variability (20\,\%). However there is a systematic preference for objects in 2MASS to be brighter than the UGPS counterpart, most likely as a result of the improved resolution in UGPS being able to distinguish between stars that appear as single objects in 2MASS.

We cross-matched our list of variable sources (with more than 1\,mag of variability) with the list from UGPS (\citet{contreras2014} and Lucas et al. 2017; \textit{subm.}). In total, only five of our sources matched with theirs (our outflow number and the corresponding MHO number, followed by the Variable ID number from Lucas et al. 2017): \\ \\
OF\,020 (MHO\,4016): V110 (separation 0.27\arcsec) \\
OF\,031 (MHO\,3929): V93 (separation 0.06\arcsec) \\
OF\,090 (MHO\,3872): V80 (separation 0.055\arcsec) \\
OF\,114 (MHO\,3840): V40 (separation 0.08\arcsec) \\
OF\,324 (MHO\,4002): V109 (separation 0.006\arcsec) \\

Of the remaining stars, most were not included in the Lucas et al. catalogue due to the stars having K\,$>$\,16, being saturated or being otherwise flagged as not having perfectly reliable photometry in one or both epochs in UGPS. Three of our variable sources are counted twice due to being source candidates for multi-polar outflows, and some of the sources show circumstellar nebulosity and hence are not point sources (excluding them from the Lucas et al. catalogue). It should be noted that some of our sources may not be genuine variables since we did not exclude those with nebulosity, but we include them for completeness. 

\subsection{MHO properties}

\subsubsection{Outflow orientation} \label{sec:orientation}

\begin{figure}
  \center
    \includegraphics[width=\columnwidth]{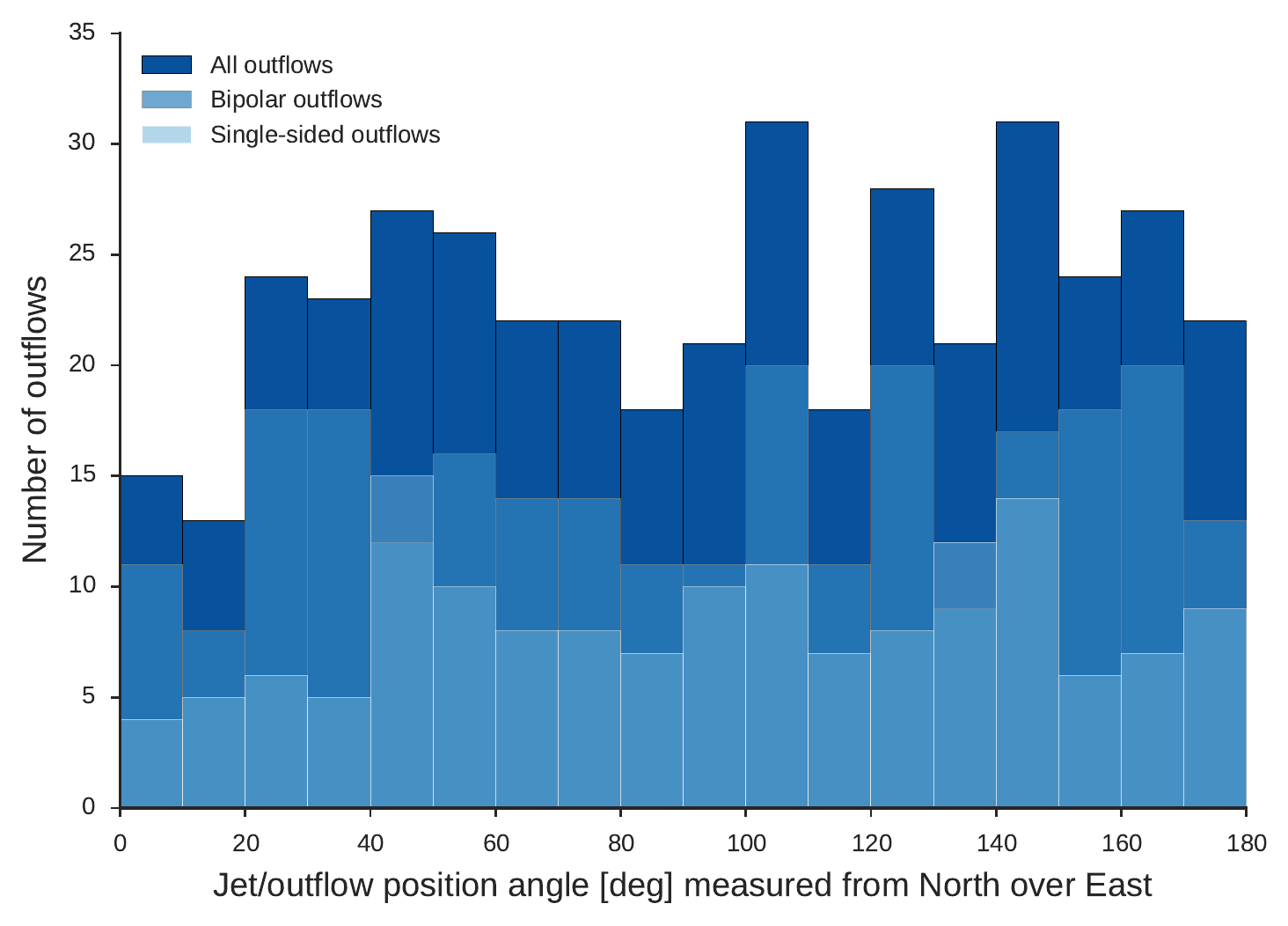} 
    \caption{The outflow position angles, measured from North towards East. For the bipolar distribution, we use the average position angle of each lobe pair. The bipolar and single-sided distributions are each superposed onto the total distribution.} 
    \label{fig:orientation}
\end{figure}

In Fig.\,\ref{fig:orientation} we show the distribution of position angles for all the bipolar and single-sided outflow lobes as measured from North toward East. Where the measurement results in an angle greater than 180\dg, we subtract 180\dg\ so that all angles are between 0\dg\ and 180\dg. In the case of the bipolar outflows we measure the properties of each outflow lobe separately, and hence we use the average position angle between each lobe pair to represent the orientation of the outflow axis.

In order to investigate whether there is any preferential orientation in the position angles of our outflows, we created a homogeneous test distribution between 0\,--\,180\dg. We then performed two-sample KS-tests of the outflow lobe position angles against this homogeneous distribution, with a null hypothesis that they are drawn from the same sample. We find that the bipolar outflows are homogeneously distributed ($p =$\,97.5\,\%), but that the test is inconclusive with regard to the single-sided outflows and the total (bipolar and single-sided together). We tested the single-sided and bipolar distributions against each other, but the test is also inconclusive here ($p =$\,56.7\,\%). This is not what was found in Auriga and Cassiopeia, where the test was decisively in favour of the distributions all being homogeneous. 

This being the case, we do note some apparently preferential orientation with respect to local dust lanes/filaments in the regions around e.g. DR\,21 and W\,75\,N, a result that was originally noted in \citet{davis2007}. This may partly explain the apparent ambivalence of the KS-test results. Another possible explanation may be that in Cygnus-X, we find that the bipolar outflows may not be perfectly straight. In Auriga and Cassiopeia, most were more-or-less straight, with a median angle difference between the bipolar lobes ($\Delta\,\theta$) of $\sim$\,5\dg\ and only 8 outflows (14\,\% of the bipolar flows) with $\Delta\,\theta>$\,10\dg. In Cygnus-X, although the median $\Delta\,\theta$ is 5.5\dg, there are 75 outflows (28\,\% of the bipolar flows) with $\Delta\,\theta>$\,10\dg. The reason for this large fraction of ``bent'' outflows is unclear. It may be a bias resulting from the method we used to measure the position angle, i.e., measuring to the tip of the furthest knot, rather than through the centre of the outflow axis. However only 17 (23\,\%) of the outflows with $\Delta\,\theta>$\,10\dg\ show signs of precession, so this does not account for the remaining objects. This may therefore be a result of the dustier environment in Cygnus-X deflecting the outflows.

Of the total of 572 outflows, 50 (9\,\%) exist in multi-outflow systems, typically forming X-shaped crosses of two or more bipolar flows. Four outflows form a complex structure with one long, straight outflow OF\,097 (MHO\,3879) crossed by three more, OF\,096 (MHO\,3880), OF\,098 (MHO\,3881), and OF\,099 (MHO\,3882). Six outflows form two triplet systems; OF\,247 (MHO\,3466), OF\,248 (MHO\,3467), and OF\,249 (MHO\,3468) form the first triplet and OF\,275 (MHO\,3846), OF\,276 (MHO\,3847), and OF\,586 (MHO\,3848) form the second. The remainder are quadrupolar, X-shaped outflows (although five of these have only one lobe visible). In Auriga and Cassiopeia, 10\,\% of the outflows exist in such multi-polar systems, all of them being quadrupolar, X-shaped flows (F16). It is unclear whether these are wide-binary (or higher order multiple) systems as found by \citet{lee2016} in Perseus, or simply the result of a projection effect, i.e., outflows originating from sources along the line of sight that are superposed onto each other, giving the impression of being multi-polar. One interesting point to note about outflows in these multi-polar systems is that 42\,\% of them show signs of precession, compared to only 22\,\% of the rest of the population, favouring the binarity explanation. In Auriga and Cassiopeia, the only outflows showing precession were found in multi-polar systems. Further observations would be required to properly investigate this phenomenon, although we note that of the 17 pairs of outflows forming X-shaped systems, 7 (41\,\%) are preferentially anti-aligned, i.e. with angle differences of 70\dg\,--\,90\dg\ between the outflows, with the remainder being randomly oriented (between 20\dg\ and 70\dg). None were found to be tightly aligned (less than 20\dg\ between the outflows). This is in agreement with the findings of \citet{lee2016}.

We also have ten outflows in Cygnus-X forming close parallel pairs with less than 20\arcsec\ separation between the driving sources; OF\,018 and OF\,019 (MHO\,979 and MHO\,978), OF\,041 and OF\,581 (both MHO\,3426), OF\,138 and OF\,139 (MHO\,3861 and MHO\,3860), and finally, OF\,172 and OF\,173 (MHO\,3556 and MHO\,3555). 

\subsubsection{Outflow lengths}
\begin{figure*}
\centering 
\includegraphics[angle=0,width=0.495\linewidth]{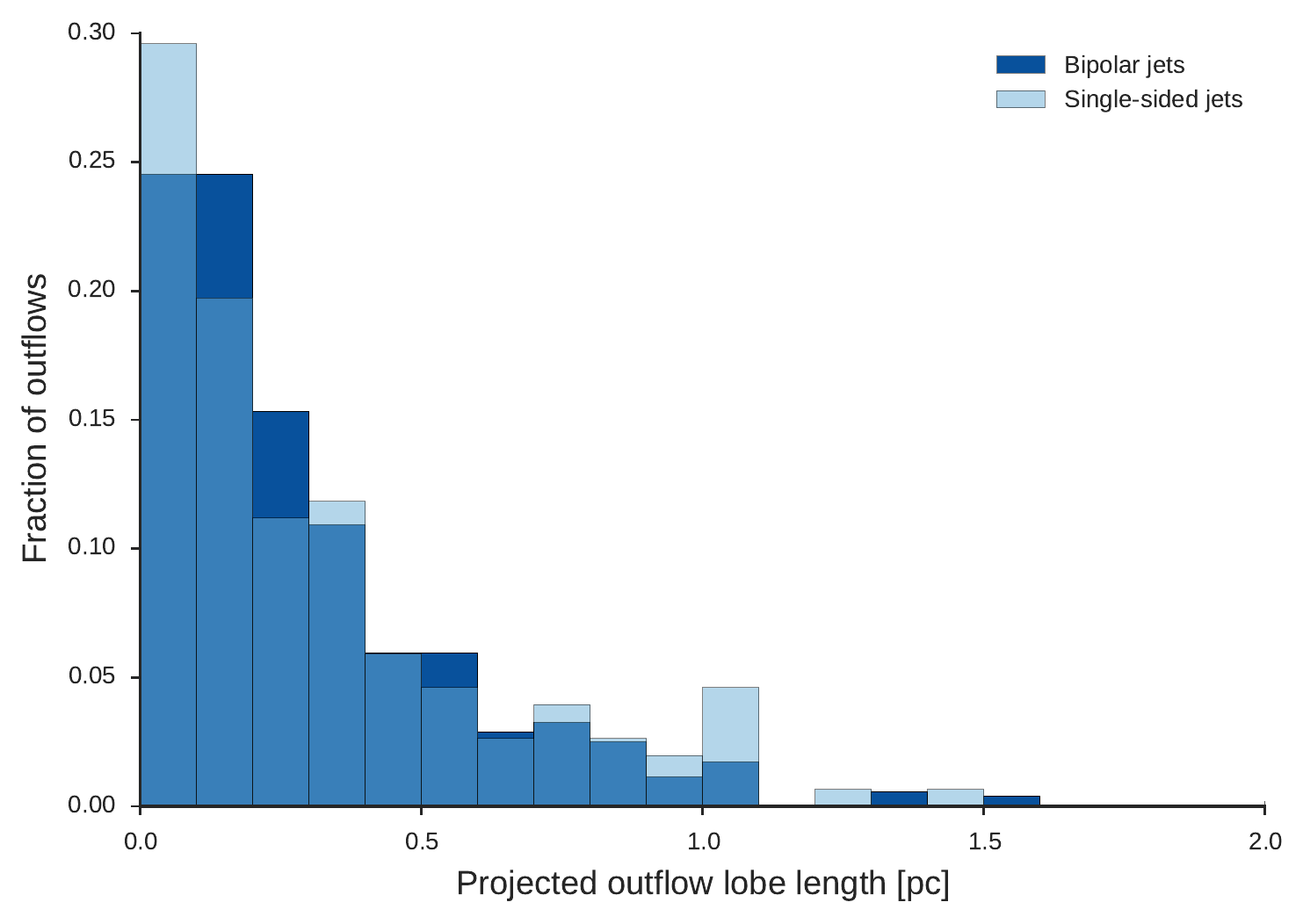} 
\includegraphics[angle=0,width=0.495\linewidth]{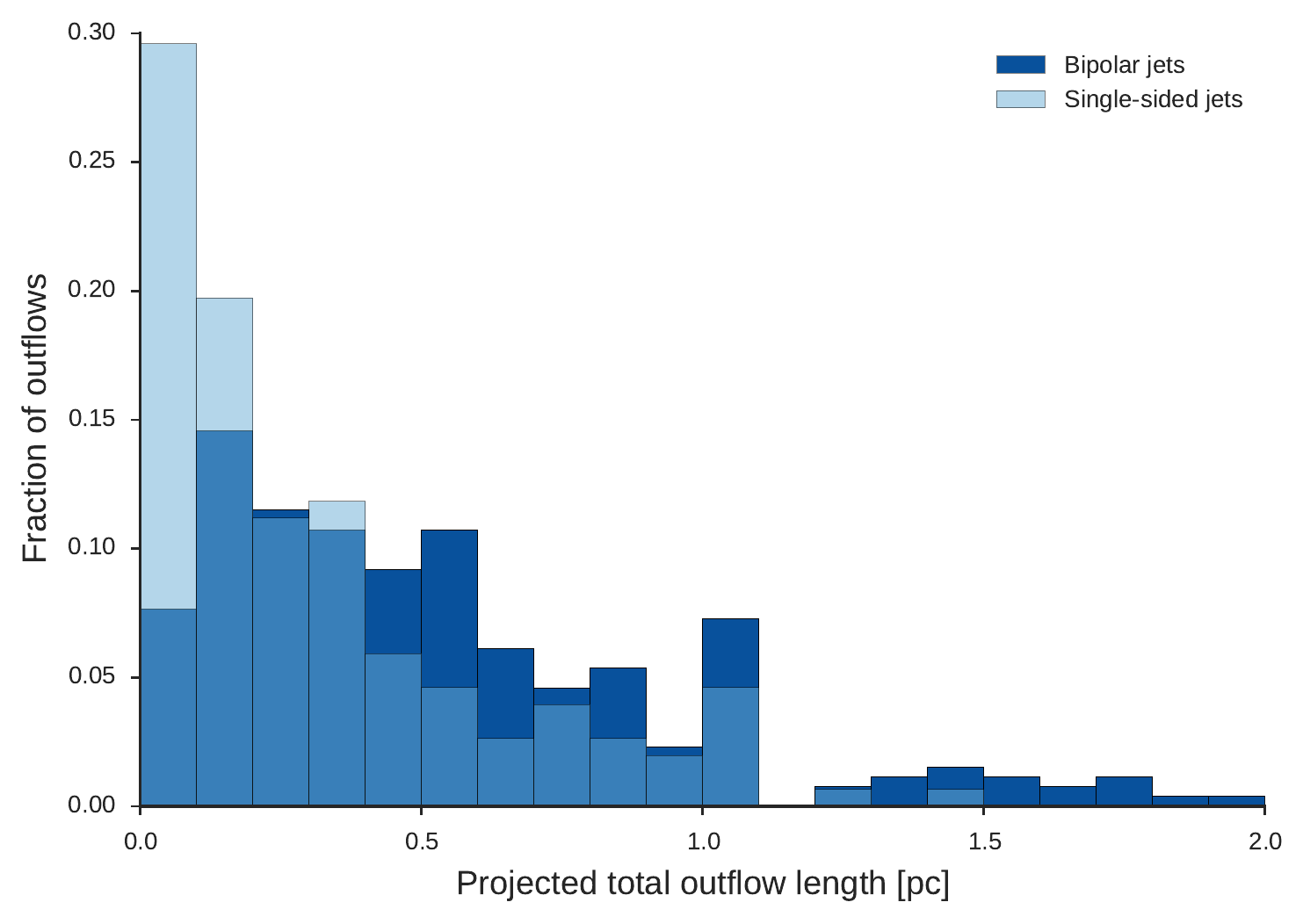} 
\\ 
\includegraphics[angle=0,width=0.495\linewidth,keepaspectratio]{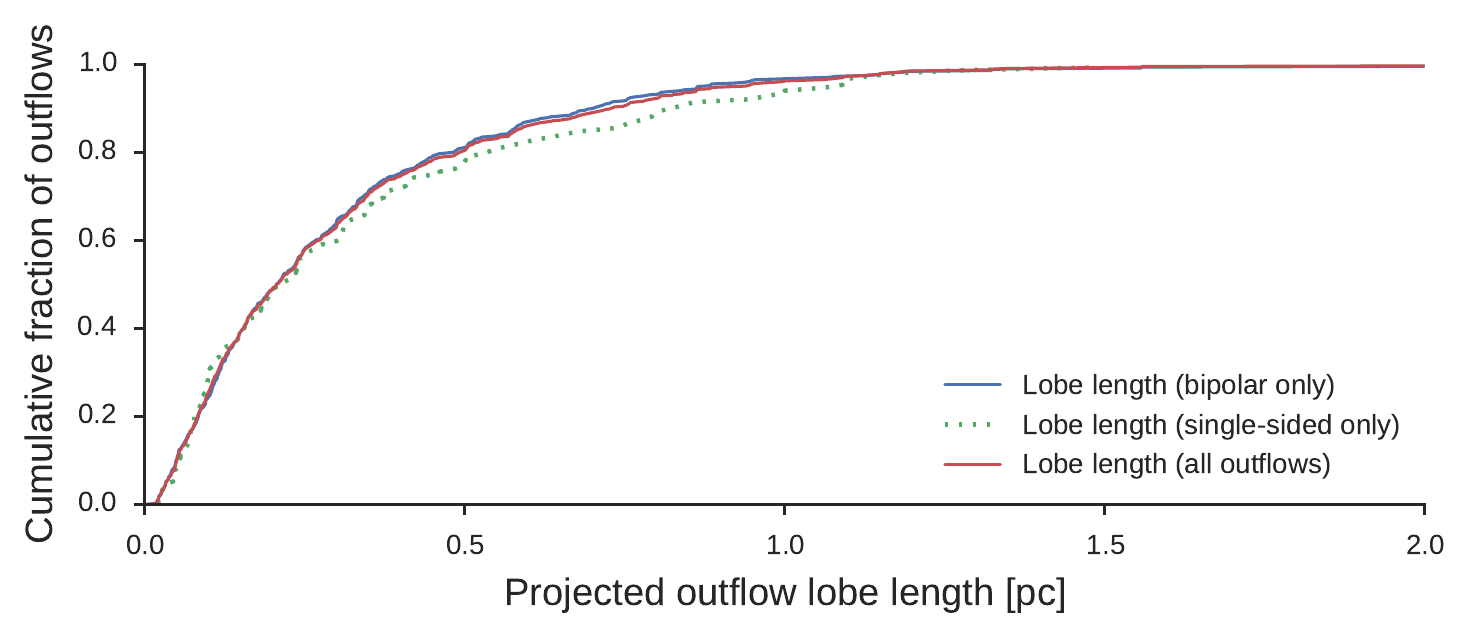} 
\includegraphics[angle=0,width=0.495\linewidth,keepaspectratio]{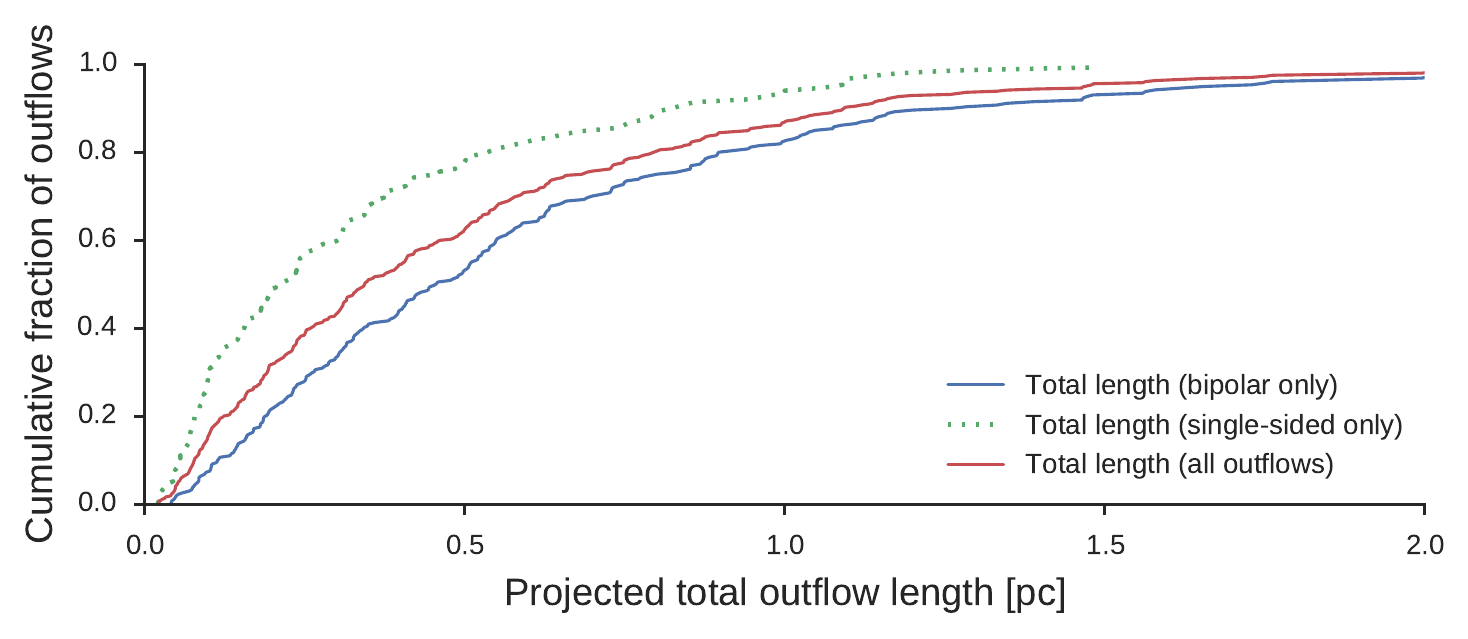}
\caption{Bipolar vs. single-sided (top row) distributions and (bottom row) corresponding cumulative distribution functions for: (left) projected lobe lengths and (right) projected total outflow lengths. The histograms are normalised to the same total number of objects, and in all four panels the x-axis has been truncated to 2\,pc. This excludes two outliers from the left-hand panel and seven from the right-hand panel. \label{fig:ld_single_bipolar}}   
\end{figure*}

Traditionally, outflow lengths are measured from end-to-end (``total length''), irrespective of whether one or both lobes are visible. We measured the length of each outflow lobe separately, from the coordinates of the most likely driving source, to the furthest tip of the most distant knot from that source (``lobe length''). In this case, we found 41 (16\,\%) parsec-scale bipolar outflows (i.e., with a total length over 1\,pc at our assumed distance of 1.4\,kpc), and the median total length of the bipolar flows is 0.45\,pc. As was found in Auriga and Cassiopeia (F16), we note that there is a distinct difference in the length distributions produced by these measurements, and hence the way that the outflow lengths are measured is important. We demonstrate this clearly in Fig.\,\ref{fig:ld_single_bipolar}.

In the top row we show the projected lobe length distributions. On the left, we show the results of measuring the lobe lengths independently, and on the right the total outflow lengths (end-to-end), in each case splitting the distribution into bipolar and single-sided outflows. Since the bipolar outflows each have two lobe length measurements associated with them, we normalised all distributions to the same total number of objects. Note that the ``tail'' of very long outflows continues up to just under 4\,pc in the right-hand panel due to some extremely long outflows (such as the W\,75\,N main outflow), so the x-axis has been truncated. This excludes two outflows from the left-hand panel and  seven from the right-hand panel. Some of the outlying, very long outflows are discussed further in Sect. \ref{sec:case_studies}. It is likely that these measurements represent lower limits to the outflow lengths since we only measure the parts of the outflow interacting with \htwo\, and the outflows will lose their molecular tracers to the UV-rich ISM once they emerge from their natal clouds. An example of this is presented in Sect. \ref{sec:OF180}.  

We KS-tested the bipolar and single-sided distributions against each other for both of these scenarios. For the lobe lengths, the KS-test is inconclusive ($p = $\,62\,\%), but it very clearly rejects the total outflow length distributions being from the same parent sample ($p = $\,4\,$\times 10^{-8}$\,\%). This is represented graphically in the bottom row of Fig.\,\ref{fig:ld_single_bipolar}, where we show the cumulative distribution functions (CDFs) that correspond to the top row. The dashed line represents the single-sided lobe distribution in each panel, which provides a useful point of reference since it doesn't change whether one measures the lobe length or total length. This shows clearly that the distributions of bipolar vs. single-sided lobe lengths are more similar, to each other and to the sample as a whole, than the total length distributions which are all poorly correlated with each other. Together, these results suggest that measuring outflow lengths in the traditional way (from end-to-end), or at least mixing the bipolar and single-sided outflows when measuring the total lengths, should be avoided. Therefore, where we refer to ``length'' or ``length distribution'' in the remainder of this paper, we intend to include all single-sided and bipolar lobe lengths, and not the total lengths.

\subsubsection{Outflow asymmetry}
\begin{figure*}
 \center
   \includegraphics[width=0.495\linewidth]{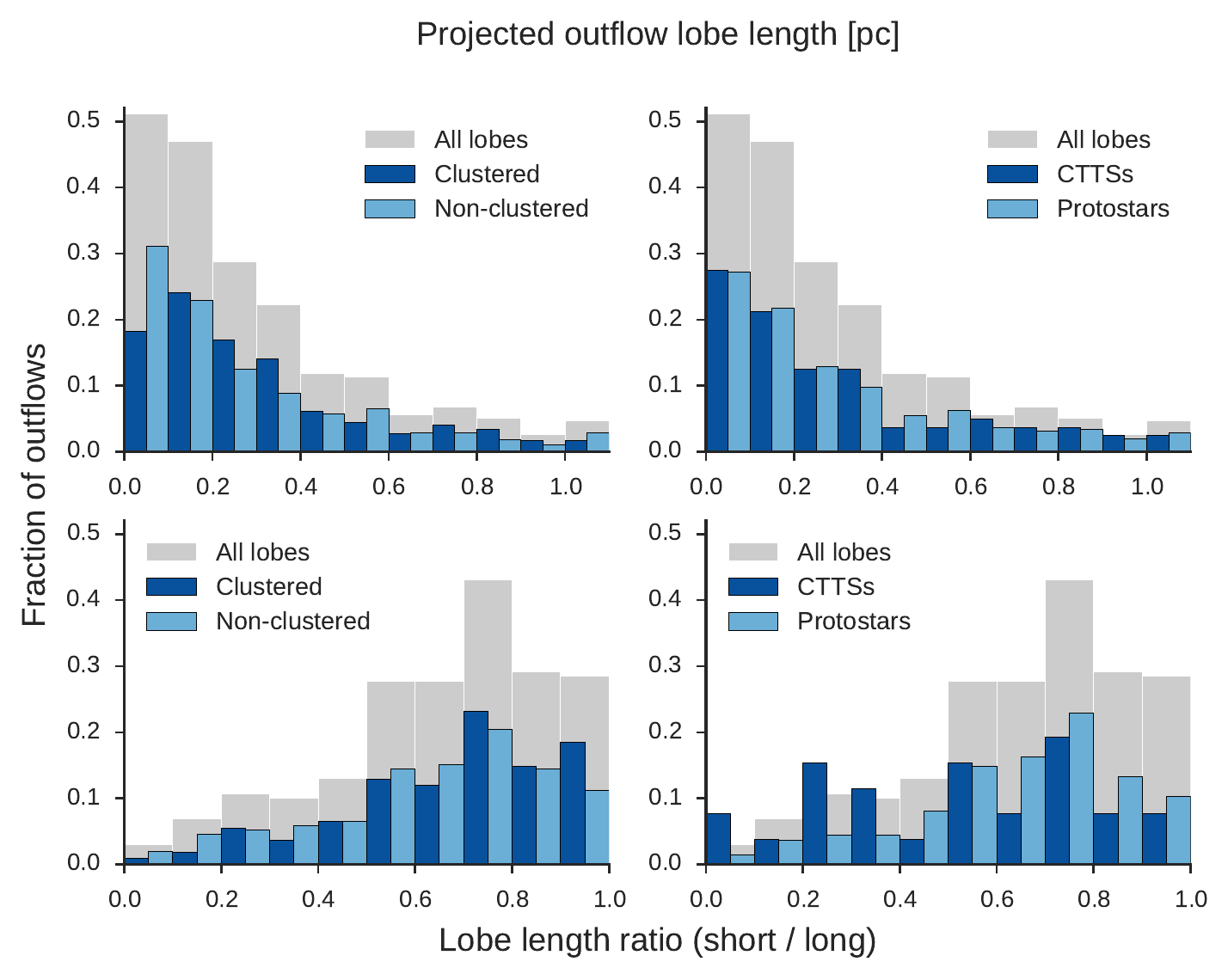} \hfill 
   \includegraphics[width=0.495\linewidth]{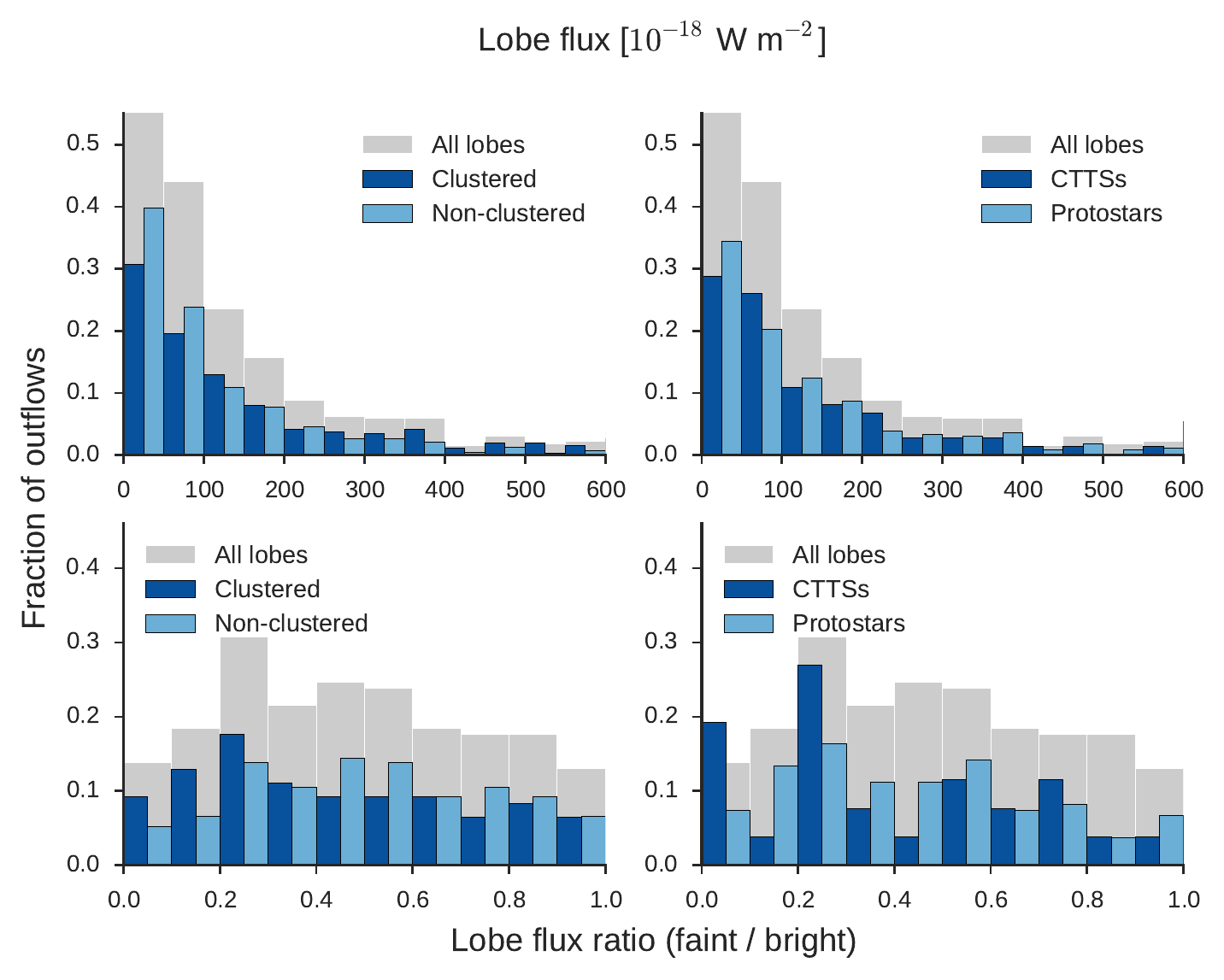} 
   \caption{\label{fig:panel_length} {\bf Left-hand panels:} Lobe length distributions (top pair) and lobe length ratio distributions (bottom pair). {\bf Right-hand panels:} Lobe flux distributions (top pair) and lobe flux ratio distributions (bottom pair). In both sets of panels the distributions are split on the left for clustered and non-clustered environments and on the right for protostellar and CTTS driving sources.  All histograms are normalised to the same number of objects. For reference, we show the distribution of ``all'' lobe lengths in the background, normalised and then stretched by a factor of two for visibility. } 
\end{figure*}

In order to investigate the degree of symmetry between pairs of outflow lobes, we calculated the ratio between the lengths of the shorter lobes over the longer ones for all bipolar outflows. A perfectly symmetrical outflow would have a length ratio of 1, and all values are between 0 and 1. We find that the typical outflow is slightly asymmetrical, with a median length ratio of 0.7 and the length ratios are mostly homogeneously distributed between 0.5 and 1.0. We show this in the left-hand panels of Fig.\,\ref{fig:panel_length}. We pair the lobe length distributions (top left pair of panels) with the length ratio distributions (bottom left pair of panels), split for clustered/non-clustered environment, and protostellar\,/\,CTTS driving sources. We overlaid these onto the parent distributions (``all'' lobes) shown in grey to give a point of reference. Whilst all the histograms are normalised to the same total number of objects, the background is stretched by a factor of two to improve visibility.

KS-testing for each pair of sub-samples in Fig.\,\ref{fig:panel_length} showed that the outflow lengths in clustered and non-clustered environments are not from the same distribution ($p =$\,0.07\,\%), but was inconclusive for the outflows from protostars vs those from CTTSs. The very shortest outflows tend to come from non-clustered environments. The situation is reversed for the lobe length ratios, where the test was inconclusive for the clustered and non-clustered environments, but showed that the flows driven by protostars and CTTSs are clearly from different distributions ($p =$\,0.8\,\%). Although there are far fewer highly asymmetrical outflows, they tend to be preferentially driven by older CTTSs. It is not clear if this is a selection effect caused by the fact that we only have WISE detections for around half of our outflows, or if it reflects a real evolutionary trend where outflows begin as symmetrical and become less so over time due to interactions with an inhomogeneous interstellar medium. 

\subsubsection{Knot properties: knot fluxes and knot gaps}
\begin{figure}
 \center
   \includegraphics[width=\columnwidth]{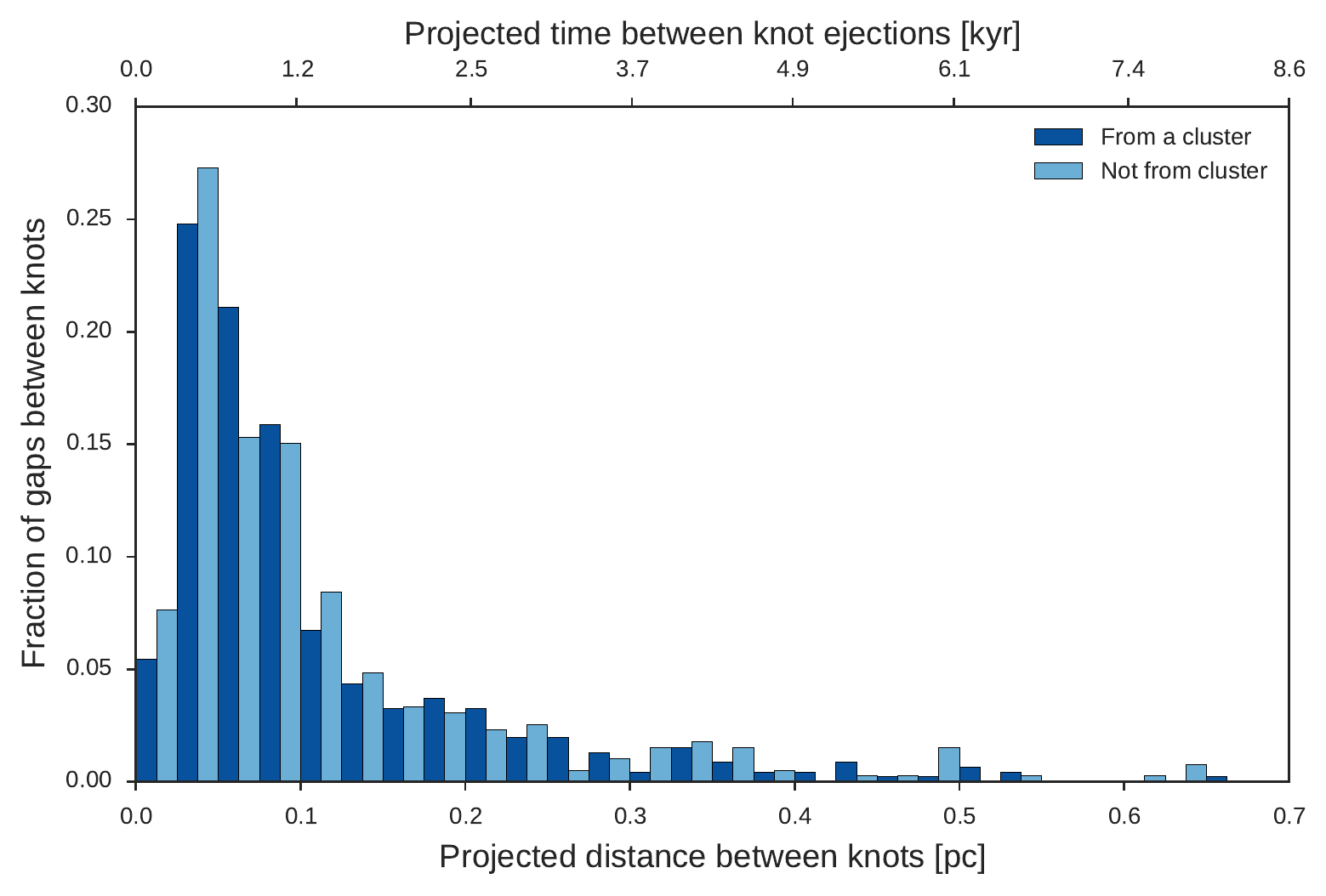} 
   \caption{The distribution of gaps between subsequent, individual outflow knots, split by whether or not the outflow that the knots are part of originates from a cluster. The axis is truncated at 0.7\,pc, with five objects not shown. Both histograms are normalised to the same total number of objects.} 
   \label{fig:knotgaps_tiny}
\end{figure}

We measured the gaps between each subsequent knot of \htwo\ emission for the bipolar and single-sided outflow lobes, and we show the distribution of these knot gaps in Fig.\,\ref{fig:knotgaps_tiny}. The gaps between each pair of knots can either reveal the locations of denser parts of the local environment, or represent the timescales between mass accretion/ejection events, or both. The median distance between the knots is 0.07\,pc, and the mean is 0.12\,pc. This corresponds to $\sim$\,0.9\,--\,1.4\,kyr between each knot, if we assume a projected transversal speed of 80\,\kms\ and we assume a constant distance of 1.4\,kpc toward Cygnus-X. The typical range of lengths between knot gaps is between 0.025\,--\,0.1\,pc, which corresponds to 0.3\,--\,1.2\,kyr. The time between ejection events was 1\,--\,3\,kyr in Auriga and Cassiopeia (F16) and 1\,--\,2\,kyr in Serpens and Aquila (IF12b).  

Assuming a distance of 1.4\,kpc toward Cygnus-X, the typical projected lobe length is up to $\sim$\,0.5\,pc (for parsec-scale outflows). This corresponds to a dynamical timescale range of 0.5\,--\,6\,kyr, which limits the occurrence of major mass accretion events to being typically every kyr or so for the YSOs driving the outflows. This rules out FU-Ori type objects as possible driving sources for the majority of our outflows, given that they erupt on timescales of 5\,--\,50\,kyr \citep{scholtz2013}, and EX-Ors \citep{reipurth2010}, which erupt on a semi-regular basis every 1\,--\,10\,yr \citep{audard2014}. Recently, a new intermediate class of eruptive variable (MNors) has been suggested in \citet{contreras2017b} following the discovery of a population of YSOs with an outburst frequency of $\sim$100\,yr, but this is still lower than the timescales suggested by the typical gaps between outflow knots.   

As shown in Fig.\,\ref{fig:knotgaps_tiny}, we split the distribution of the gaps between knots according to whether or not the outflow driving source is situated in a cluster. KS-testing is inconclusive about whether or not the clustered and non-clustered sub-samples are drawn from the same distribution ($p = $\,58\,\%), and was similarly inconclusive when we tested the outflows driven by protostars and CTTSs against each other ($p = $\,78\,\%). However, the median gap lengths are the same for both the clustered and non-clustered outflows (0.072\,pc and 0.076\,pc respectively) and for outflows driven by protostars and CTTSs (0.075\,pc and 0.074\,pc respectively). One might expect that the protostars would have mass accretion bursts more frequently \citep{VB2006}, and that we therefore might measure shorter gaps between each outflow knot than from the older driving sources, but we do not see such a difference. We might also expect to see a significant difference in the dynamical timescales between the major knots from clustered and non-clustered environments, but we do not see any such difference here either.

\subsubsection{Outflow fluxes}
We find that the typical bipolar outflow is faint in \htwo, with a median total flux of 18\,\Wm\ ($\sim$\,1.1\,$\times$\,$10^{-3}$\,\Lsolar). The average total flux for bipolar flows, however, is 111\,\Wm\ (6.8\,$\times$\,$10^{-3}$\,\Lsolar) as a result of extreme outliers such as the DR\,21 and W\,75\,N outflows. 

\begin{figure*}
 \center
	\includegraphics[width=0.495\linewidth]{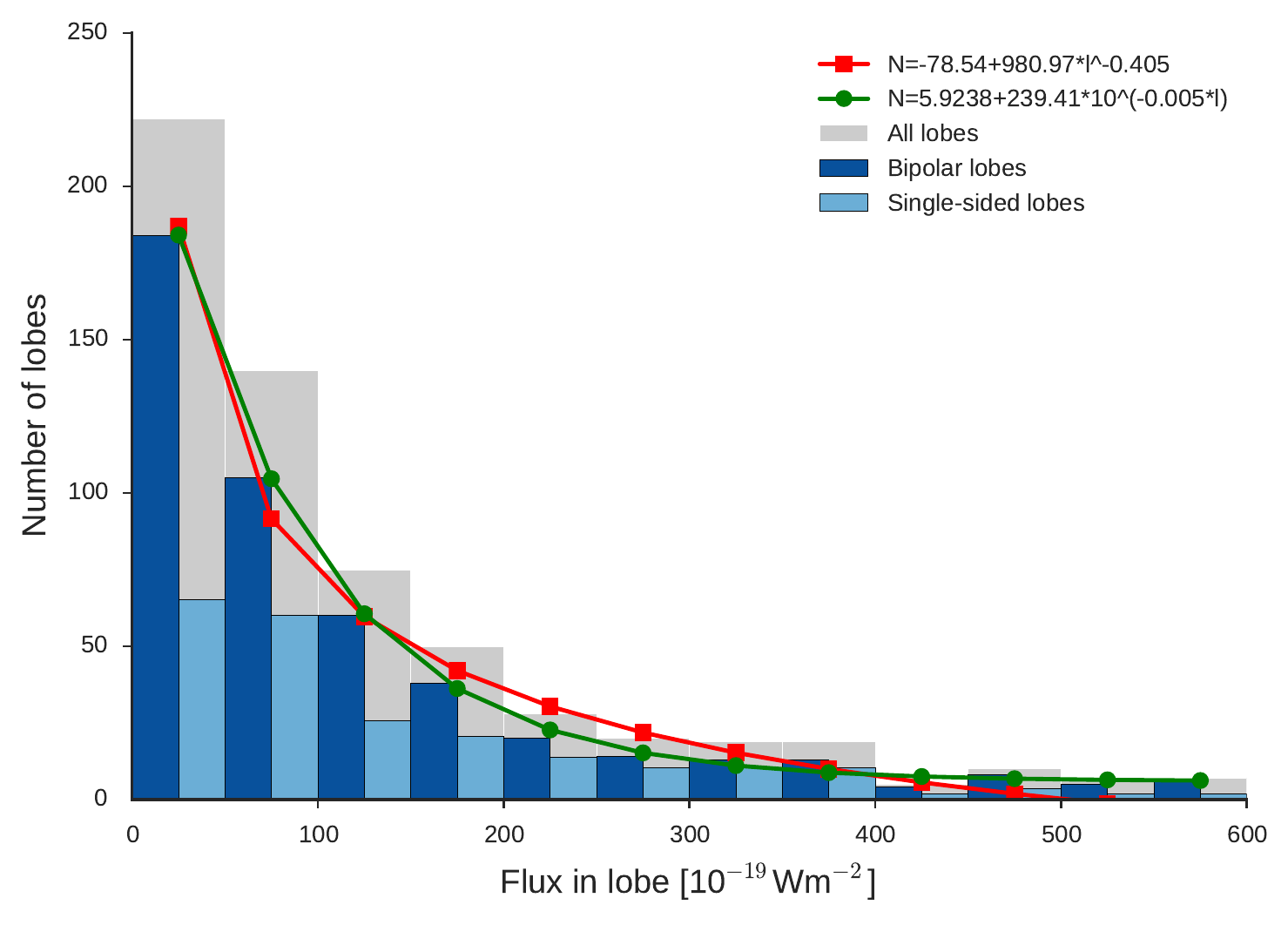} \hfill 
	\includegraphics[width=0.495\linewidth]{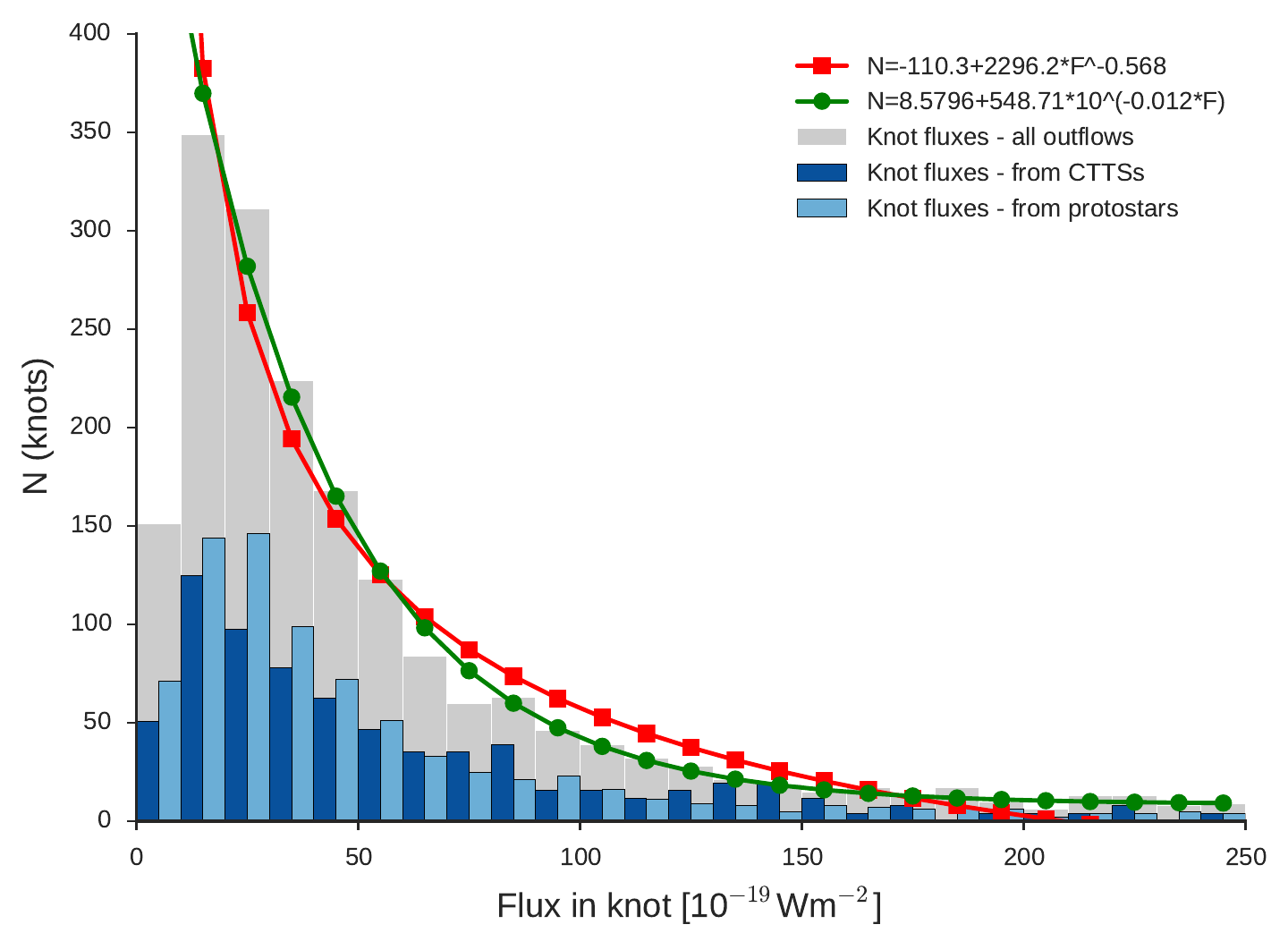} 
   \caption{(Left) The distribution of fluxes in the outflow lobes, splitting the bipolar and single-sided distributions. The total distribution of ``all'' outflow lobes is shown in the background to provide a point of reference. Due to a large number of high-value outliers, the $x$-axis has been truncated so the 60 brightest outflows are not shown on this graph. (Right) The distribution of fluxes in every knot (including only knots from bipolar and single sided outflows) is shown in grey. In both panels, the distributions have been scaled to the same total number of objects. The red line shows the powerlaw fits, whilst the green shows the exponential fits.
   \label{fig:fluxes}}
\end{figure*}

In the left-hand panel of Fig.\,\ref{fig:fluxes} we plot the distribution of the total fluxes from every outflow lobe (grey), with the distribution split for bipolar (dark blue) and single-sided (light blue) lobes over the top. We find that the data are better fit by an exponential ($RMS = 3.8$) than by a powerlaw ($RMS = 9.1$). The distribution starts to deviate from a powerlaw above a flux of about 35\,\Wm, which is comparable to the result found in Auriga and Cassiopeia (F16), where the powerlaw starts to deviate above 30\,\Wm. The overall lobe flux distribution follows $N \propto F^{-0.4}$, which is shallower than the $N \propto F^{-0.5}$ found in Auriga and Cassiopeia, and is outside the $N \propto F^{-0.5...-0.7}$ found in Serpens and Aquila by IF12b. 

\begin{figure}
	\centering 
		\includegraphics[angle=0,width=\columnwidth]{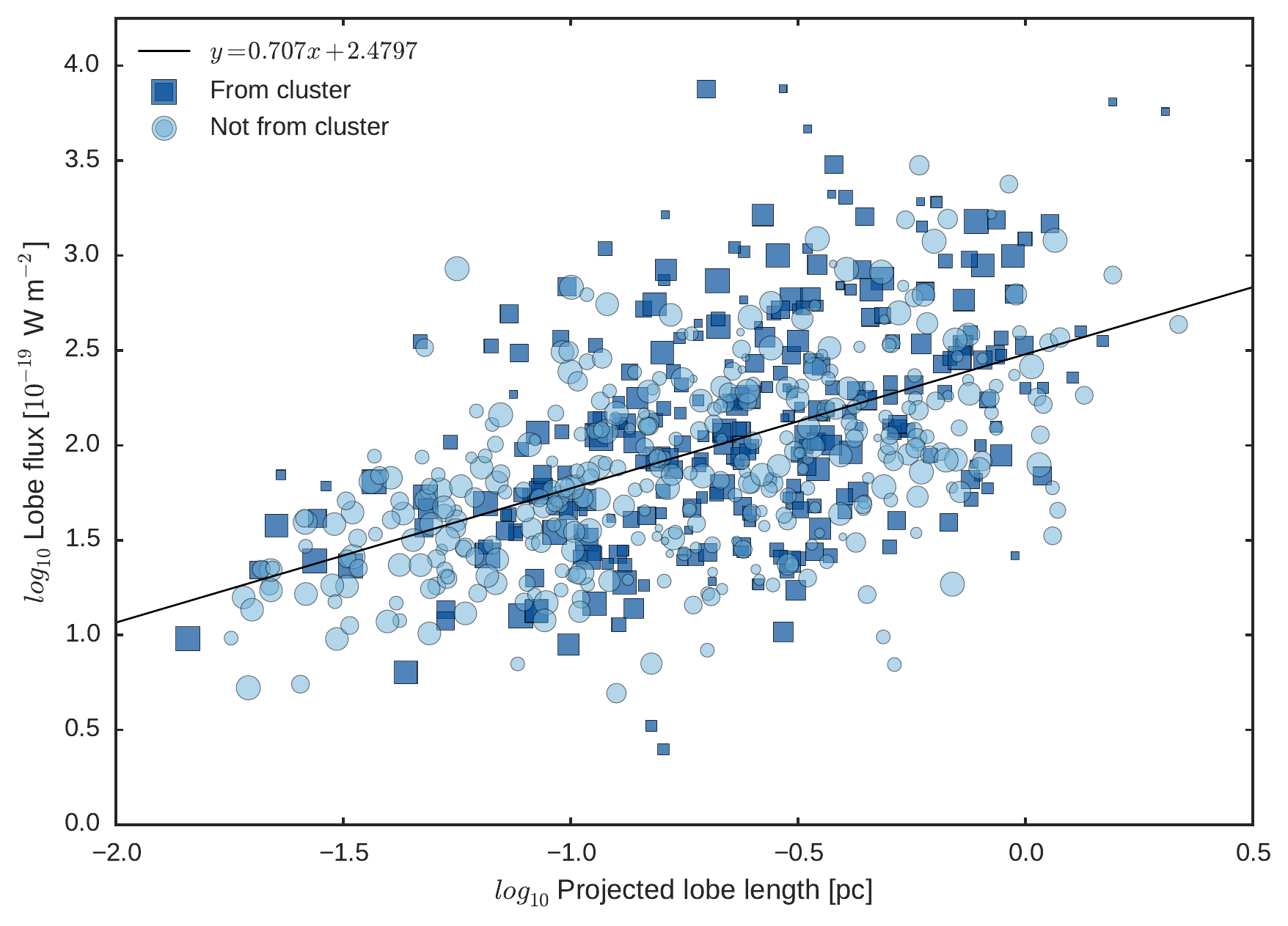} 
		\caption{A log-log plot of the length of each lobe vs. the flux in that lobe, which includes bipolar and single-sided outflows. There is a weak linear correlation between the length and brightness ($r =$\,0.52). The size of the points correspond to our confidence in the selected source (large points are more likely to be correctly identified).}
		\label{fig:len_vs_flux}
\end{figure}

In the right-hand panel of Fig.\,\ref{fig:fluxes} we show the distribution of fluxes in all of the knots, and also performed powerlaw and exponential fitting to this data. We find that as with the lobe fluxes, the distribution is better fit by an exponential, with the $RMS$ of the powerlaw being twice as high. The distribution of the knot fluxes, when fit by a powerlaw, follows $N \propto F^{-0.6}$ (exactly the same as was found in Auriga and Cassiopeia (F16)). In both cases, there are many bright outliers which occur too frequently for the distributions to follow a powerlaw; both deviate from a powerlaw above 15\,\Wm. The $RMS$ of the powerlaw fit is around twice as high as the exponential fit for all cases. We should note that in the left-hand panel of Fig.\,\ref{fig:fluxes}, there are 60 outflow lobes (mostly from very active star-forming regions) which have a flux much greater than 60\,\Wm\ and are not shown, and in the right-hand panel there are 37 knots not shown beyond 25\,\Wm. Note that we exclude the single knots from both panels of this figure, although we did measure their fluxes. When we split the distribution of fluxes into bipolar\,/\,single-sided outflows and single knots, we found no statistical difference in the distributions of fluxes from knots in single-sided outflows and from single knots ($p =$\,99.45\,\%), but the knots from bipolar flows have a different distribution from them both ($p =$\,4.81\,\% and $p =$\,0.14\,\%, for single-sided flows and single knots respectively).  

In the right-hand set of panels in Fig.\,\ref{fig:panel_length} we show the lobe flux (top right pair) and lobe flux ratio (bottom right pair) distributions in the same way as we did with the lobe length distributions. We find that the median flux ratio of the faint lobe over the bright one is 0.46 suggesting that in the typical bipolar outflow, one lobe is twice as bright as the other. Again, we performed KS-testing of each of these distributions against each other. The flux distributions from clustered and non-clustered environments are most likely not drawn from the same sample ($p =$\,0.08\,\%), and results are inconclusive for the flux ratios. The very faintest outflows are slightly preferentially driven by isolated outflows. The KS-test was inconclusive when we tested the distributions of protostellar- and CTTS-driven outflows against each other for the lobe fluxes, but was clear for the flux ratios that they are drawn from different samples ($p =$\,5.1\,\%). The greatest asymmetry occurs in outflows with CTTSs as driving sources. Since we saw the same result with the length ratios, this implies there may be some correlation between the length and flux in each lobe.

In Fig.\,\ref{fig:len_vs_flux} we plotted the $log_{10}$\,lobe length against the $log_{10}$\,lobe flux. We find a weak positive correlation between the length of a lobe and its brightness with correlation coefficient $r =$\,0.52 and a confidence level of $\sim$\,26\,\%. Once again, splitting the sample by clustered vs. non-clustered environments or by evolutionary stage seems not to make any difference. Note the two lobes of the DR\,21 main outflow are considerably brighter than all other outflows and fall outside of the axis limits.

\section{Interesting Individual Outflows} \label{sec:case_studies}
Here we discuss a selection of outflows in more detail. For the full list of outflows images and brief descriptions on each of them, please refer to the online journal and Appendix\,\ref{sec:appx_images} for an example of the format and content. Note that where we discuss the outflow \htwo\ luminosities, these have not been corrected for extinction and only include the contributions from the 1\,--\,0\,S(1) line of \htwo. 

\subsection{Some well-known outflows}

\subsubsection*{DR\,21: OF\,570 (MHO\,898, MHO\,899)} 
The main outflow of DR\,21 has, by a long margin, the highest 1\,--\,0\,S(1) \htwo\ flux of any object in our sample. This outflow alone contains 42\,\% of the total \htwo\ flux presented in this catalogue. The lobe extending toward the north-east is red-shifted, with the blue-shifted lobe extending toward the south-west \citep{garden1991}. We find that the lobes are almost symmetrical in length ($R_L =$\,0.89) and flux ($R_F =$\,0.91), with the red-shifted lobe being the slightly shorter and fainter one.

There has historically been some of disagreement about the correct distance toward DR\,21, with typical estimates ranging between 1.5\,kpc and 3\,kpc (e.g., \citet{genzel1977} - 3\,kpc; \citet{fischer1985} - 2\,kpc; \citet{odenwald1993} - 2\,kpc, \citet{schneider2006} - 1.7\,kpc).\, \citet{davis2007} used a distance toward DR\,21, W\,75\,N and L\,906\,E of 3\,kpc, but \citet{rygl2012} determined the distance toward DR\,21 of 1.5\,kpc using maser parallax. Using the Rygl distance, the total 1\,--\,0\,S(1) \htwo\ luminosity is 1.1\,\Lsolar, and the projected total length is 2.7\,pc, making this also one of the longest outflows in the sample. However, if we assume the distance to DR\,21 is closer to the 3\,kpc value used by \citet{davis2007} then the projected total length increases to 5.4\,pc and the total 1\,--\,0\,S(1) \htwo\ luminosity to 4.3\,\Lsolar.

\subsubsection*{AFGL\,2591: OF\,213 (HH\,166) (MHO\,952, MHO\,953)}
The second-highest 1\,--\,0\,S(1) \htwo\ flux comes from OF\,213, driven by the well-studied object AFGL\,2591--VLA3 \citep{johnston2013}. The outflow has been detected at various wavelengths and was first named as a Herbig-Haro object (HH\,166) in \citet{poetzel1992}. Originally, the distance to this object was estimated at 1\,--\,2\,kpc (from e.g. \citet{poetzel1990,poetzel1992}, \citet{trinidad2003}, \citet{vandertak2005}), but \citet{rygl2012} measured 3.33\,kpc\,$\pm$\,0.11. At a distance of 3.33\,kpc, \citet{johnston2013} determine the luminosity of the driving source as 2.3\,$\times$\,$10^{5}$\,\Lsolar. We determine the total 1\,--\,0\,S(1) \htwo\ luminosity of the outflow to be 0.42\,\Lsolar\ at the same distance, with a total length of 1.47\,pc. In this case, the blue-shifted lobe (extending to the west) is shorter than the red-shifted lobe, and 1.6 times brighter. 

\subsubsection*{W\,75\,N: OF\,510 (MHO\,828, MHO\,857, MHO\,855, MHO\,856, MHO\,854)} 
The third-highest 1\,--\,0\,S(1) \htwo\ flux comes from the W\,75\,N main outflow, which is also one of the longest outflows in the sample. Given the \citet{rygl2012} distance toward the outflow of 1.3\,kpc\,$\pm$\,0.07, we calculate the total length to be 3.33\,pc with a total 1\,--\,0\,S(1) \htwo\ luminosity of 0.064\,\Lsolar. 

\subsection{Highlights from the newly discovered outflows} \label{sec:newbies}

\begin{figure*}
		\centering
	   	\includegraphics[angle=0,width=1.0\linewidth]{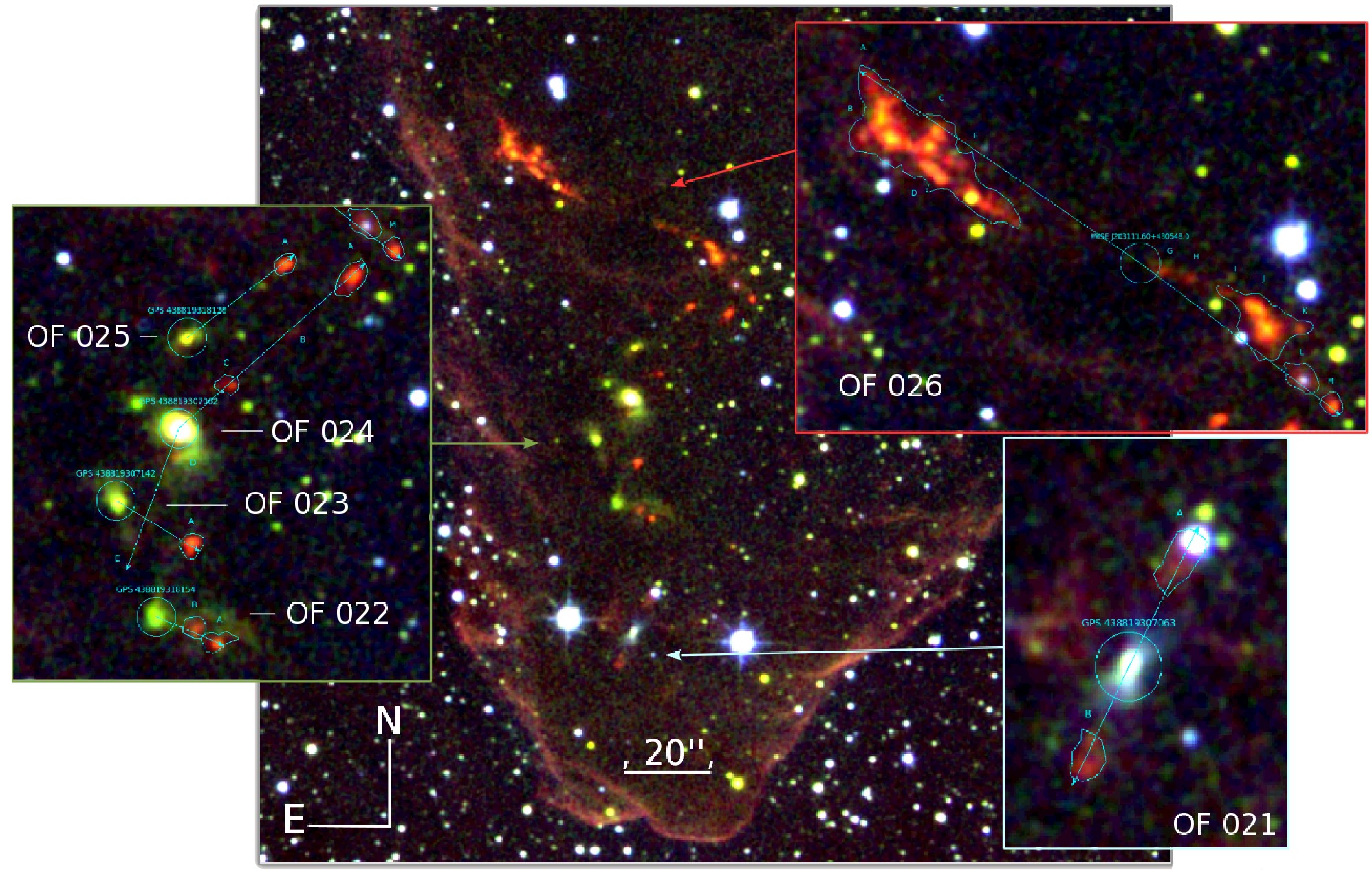}
    	\caption{
        A low-mass bright-rimmed cloud near IRAS\,20294$+$4255, which may warrant further study as a classical example of triggered star formation. 
        The cloud is illuminated from the south, suggesting compression by an O/B source. Refer to Sect. \ref{sec:newbies} for further details on this region.
        }
    	\label{fig:pretties2}
\end{figure*}

\subsubsection*{OF\,021 to OF\,026, potentially triggered star formation}
A series of six outflows are located at the tip of a bright rimmed cloud near IRAS\,20294$+$4255, shown in Fig. \ref{fig:pretties2}. This low-mass region warrants further study as a classical example of triggered star formation. The cloud itself appears to be illuminated by an O/B star to the south, given the fluorescent \htwo\ emission around the southern rim of the cloud. The southernmost outflow, OF\,021 (MHO\,3567), appears as a north-south oriented pair of emission knots either side of a source that has wings of J- and H-band reflection nebulosity along the north-south axis. The UGPS and 2MASS colours suggest that the star is less reddened than the others in this cloud, and it has no WISE detection, suggesting it is more evolved. The knots are also fainter than those in the other outflows. 

The outflows OF\,022 (MHO\,3569), OF\,023 (MHO\,3571), and OF\,024 (MHO\,3568) have a less certain assignment of knots to outflows, but all potential driving source candidates are reddened and have K-band reflection nebulae in directions either perpendicular or parallel to OF\,021. Each of these stars is a known YSO in \citet{kmh2014}; their $\alpha$ values from Spitzer, like ours (from WISE), are all positive ($\sim+1$) suggesting at least Class I objects. The two most northerly outflows, OF\,025 (MHO\,3566) and OF\,026 (MHO\,3564) both have driving sources with $\alpha\sim+2$. OF\,026 is the most northerly and is not visible at shorter wavelengths than 3.6\,\mic. It also drives a strongly precessing S-shaped outflow whose \htwo\ flux is four times higher than the emission from OF\,021 to OF\,025 combined. Given the lack of any NIR detection of the source of OF\,026 and its apparently precessing outflow, this could well be a very young binary Class\,0 source.

\subsubsection*{OF\,037 (MHO\,4044)}
V\,1219\,Cyg \citep{stecklum2013} is one of most variable driving sources in the sample, with (K$_2$\,-\,K$_1$) = -0.84\,mag and (K$_1$\,-\,K$_{\sc S}$) = -1.11\,mag (see top left, Fig. \ref{fig:f2}). We detected two \htwo\ bow shocks leading away from this star, and interestingly, evidence of shocks strong enough to produce \feii\ in the curved H-band emission (green in our image) leading away from and apparently connected to the star. It is not clear if this is reflection nebulosity around the star, but this emission also has a diffuse trail connecting it to the north-eastern bow shock which also contains H-band emission, suggesting we may be seeing a partly-illuminated cavity.  

\subsubsection*{OF\,076 (MHO\,3972)}
This outflow has a projected total length of 0.9\,pc at the assumed distance of 1.4\,kpc, and is already known in CO as [GKM2012b]\,G82.186$+$0.105 \citep{gkm2012}. In terms of length, this is one of the most symmetrical outflows in the sample with a length ratio of $R_L = 0.96$. However, the south-east lobe is twice as bright as the north-western lobe (see top right, Fig. \ref{fig:f2}).   

\subsubsection*{OF\,085 (MHO\,4005)}
A 1.2\,pc outflow with knotty bow shocks on both lobes and a deeply embedded source, likely where the outflow axis is crossed by a dusty filament. There are no detections in any of our selected point-source catalogues; the nearest is a BGPS mm detection 20\arcsec\ away. Like OF\,076, this outflow is almost completely symmetrical in length, but the north-west lobe has twice the flux of the south-eastern lobe (see centre left, Fig. \ref{fig:f2}).  

\subsubsection*{OF\,125 (MHO\,4077)}
The only FU-Ori star in the sample, V\,1057\,Cyg, is highly variable between 2MASS and UGPS (K$_1$\,-\,K$_{\sc S}$ = 3.8\,mag) but hardly at all between the two epochs of UGPS (K$_2$\,-\,K$_1$ = 0.01\,mag). Interestingly, both OF\,037 and OF\,125 are highly asymmetrical in both their lobe lengths ($R_L=0.3$ and $R_L = 0.4$ respectively) and their lobe fluxes ($R_F = 0.5$ and $R_F = 0.3$ respectively). For both outflows, the bright lobe is also the shorter of the two. 

\subsubsection{OF\,180 (HH\,570)} \label{sec:OF180}
A short, faint outflow ($\sim$0.6\,\arcmin, or 0.25\,pc). This outflow is the central counterpart of known Herbig-Haro object, HH\,570, which is $\sim$3\,\arcmin\ in length (\citet{bally2003}, where no driving source is listed). Neither of the two lobes visible in the optical (HH\,570\,S and HH\,570\,N) are detected at all in our images, but we do see curved K-band reflection nebulae either side of [RGS2011]\,J20512.91$+$440429.6 \citep{rgs2011}, which is listed as a Class I YSO and we suggest this is the most likely driving source. In Rebull et al. they use a distance toward the North America Nebula of 520\,pc. If we use this as the distance toward the outflow then the length for the \htwo\ counterpart of HH\,570 becomes only 0.09\,pc.

\subsubsection*{OF\,197 (MHO\,3878)}
A 1.3\,pc outflow with a Bolocam mm detection in the middle of a dark cloud that crosses the outflow axis, suggesting a deeply embedded source. The outflow is known in CO under [GKM2012b]\,G81.140$+$0.687 \citep{gkm2012} (see centre right, Fig. \ref{fig:f2}).

\subsubsection*{OF\,238 (MHO\,3493)}
A 1.5\,pc complex outflow, more-or-less perpendicular to a reddened filament. The bright star selected as a potential driving source is emission-line star [D75b]\,Em*\,20$-$090. This star is also listed as [MSX6C]\,G077.9280$+$00.8711 \citep{urquhart2011}, which gives an adopted distance of 1.4\,pc, and a near/far kinematic distance of 1.8\,kpc. If we use 1.8\,kpc as the distance toward the outflow the total length is increased to 1.9\,pc (see bottom left, Fig. \ref{fig:f2}).

\subsubsection*{OF\,261 (MHO\,3497)}
A 0.7\,pc outflow that appears to run along a dusty filament. The southern lobe has almost twice the length and flux of the northern lobe. The selected source is a known YSO [KMH2014]\,J202618.88$+$391955.35 \citep{kmh2014} (see bottom right, Fig. \ref{fig:f2}).

\section{Conclusions} \label{sec:conclusions}

In order to study the statistical properties of jets and outflows from young stars in the Cygnus-X star-forming region, we have analysed 42 square degrees of images taken in the narrow-band filter centred on the 2.122\,\micron\ 1\,--\,0\,S(1) molecular hydrogen line, obtained as part of the UWISH2 survey. The area covered by the data contains most of the high column density regions within 74\dg\,$< l <$\,86\dg\ and -3\dg\,$< b <$\,5\dg. 

We investigated the \htwo-K difference images as well as JK\htwo\ colour composites to identify all detectable extended \htwo\ emission-line features that can be associated with outflows from Young Stellar Objects. Properties such as the length, orientation, and flux of all identified outflows have been consistently measured. We further identified potential driving sources and measured their properties (evolutionary stage, clustered or isolated environment) for all outflows.

In total we have identified 572 outflows in the survey area, almost half of of which (261) are bipolar and for about one quarter (152) only one side was detectable. The remaining quarter (159) is comprised of \htwo\ knots without a clear association to a single identifiable outflow. Despite the fact that Cygnus-X is a well studied region, only 107 of the outflows are previously known. Thus, this work represents a more than 430\,\% increase in the number of known outflows from young stars in this region. Based on the fraction of the UWISH2 data analysed so far, we estimate that there will be about 2000 jets and outflows, with about half of these being bipolar.

For 93\,\% of the bipolar and 96\,\% of the single sided outflows we have identified a reliable driving source location. About 40\,\% of the outflows are associated with a clustered environment, while the remainder originate from isolated sources. Using WISE data we were able to determine the slope of the SED for about half of the driving source candidates, and find that 80\,\% of them are protostars while 20\,\% are CTTSs. We further study the K-band variability over several years for driving sources where we had more than one magnitude available from UGPS and/or 2MASS. About 60\,\% of the sources are variable at the 0.1\,mag level or above, 20\,\% are variable at the 0.5\,mag level or above and about 10\,\% of all sources vary by more than 1.0\,mag. 

The distribution of outflow orientations in our sample is in agreement with a random distribution. In the typical bipolar outflows the two lobes are misaligned by about 5\dg. About 10\,\% of all outflows occur in multi-outflow systems which are mostly X-shaped, with two or more outflows originating from the same apparent (binary) source. Roughly 40\,\% of outflows in these systems show signs of precession, compared to only about 20\,\% for the remainder of the population. 

The median total length of the bipolar flows is 0.45\,pc, and 41 (16\,\%) of the bipolar outflows are over 1\,pc in total length. The median total 1\,--\,0\,S(1) flux of the bipolar flows is 18\,\Wm, which corresponds to a median outflow luminosity of 1.1\,$\times$\,$10^{-3}$\,\Lsolar\ at the assumed distance of 1.4\,kpc toward all the outflows. However, we find the typical bipolar outflow to be asymmetrical, with the median ratio between the short and long lobe lengths being 0.7. Similarly, the median ratio between the faint and bright lobe fluxes is 0.5. Furthermore, there is a weak correlation between the lobe fluxes and lobe lengths, of the form $F \propto L^{0.7}$. When fit with a powerlaw, we find that the number of lobes and the flux are related via $N \propto F^{-0.4}$, which is slightly shallower than what was found in other regions. However, the full flux distribution is better fit by an exponential distribution, in agreement with results in Auriga and Cassiopeia (F16). 

The gaps between major knots in the outflow lobes have been measured for each outflow, and their distribution analysed. Typically these gaps are between 0.025\,--\,0.1\,pc. This corresponds to a time difference between the ejection of the material generating these knots of about 0.9\,--\,1.4\,kyr assuming an ejection velocity of 80\,\kms. This time gap range is at the lower end of what was found in other regions (about a factor of two shorter; F16, IF12b). 

\acknowledgements
S.V. Makin acknowledges an STFC scholarship (1482158). The United Kingdom Infra-Red Telescope is operated by the Joint Astronomy Centre on behalf of the Science and Technology Facilities Council of the U.K. The data reported here were obtained as part of the UKIRT Service Program. The authors would like to thank the anonymous reviewer, whose insightful comments have helped improve the manuscript. 

\facility{UKIRT (WFCAM)} 
\software{SAOImage DS9, Montage, Astropy, Vizier, SIMBAD}

\appendix

\clearpage 
\newpage

\section{Table of outflow and driving source properties}
\label{sec:appx_mhodata}


\startlongtable
\begin{splitdeluxetable}{ccccccccccBccccBcccccccccccc}
\tabletypesize{\scriptsize}
\tablecaption{List of outflows.\label{tab:MHO_table}}
\tablewidth{0pt}
\tablehead{
\colhead{Outflow} 		
& \colhead{MHO} 
& \colhead{RA} 		
& \colhead{DEC} 
& \multicolumn{2}{c}{Length} 		
& \multicolumn{2}{c}{PA} 		
& \multicolumn{2}{c}{Flux}	
& \colhead{Cluster} 		
& \colhead{Outflow} 
& \colhead{Source} 	
& \colhead{Confidence} 
& \colhead{J} 		
& \colhead{H} 
& \colhead{K$_1$} 		
& \colhead{K$_2$} 
& \colhead{J}
& \colhead{H}
& \colhead{K$_{\sc S}$}
& \colhead{W1}
& \colhead{W2}
& \colhead{W3}
& \colhead{W4}
& \colhead{Detected}
\\ 
\colhead{ID} 		
& \colhead{ID}
& \multicolumn{2}{c}{(J2000)}
& \colhead{Lobe 1} 		
& \colhead{Lobe 2} 
& \colhead{Lobe 1} 		
& \colhead{Lobe 2} 
& \colhead{Lobe 1}	
& \colhead{Lobe 2} 
& \colhead{} 		
& \colhead{Type} 
& \colhead{ID} 	
& \colhead{level} 
& \multicolumn{4}{c}{UGPS}
& \multicolumn{3}{c}{2MASS}
& \multicolumn{4}{c}{AllWISE} 
& \colhead{in:}	
\\
\colhead{} 		
& \colhead{}
& \colhead{[hh mm ss]}
& \colhead{[\,\dg\ \arcmin\ \arcsec\,]}
& \multicolumn{2}{c}{[deg]}
& \multicolumn{2}{c}{[deg]}
& \multicolumn{2}{c}{$10^{-19}$\,W\,m$^{-2}$}
& \colhead{yes\,/\,no} 		
& \colhead{B/S/K} 
& \colhead{} 	
& \colhead{\%} 
& \multicolumn{4}{c}{[mag]}
& \multicolumn{3}{c}{[mag]}
& \multicolumn{4}{c}{[mag]} 
& \colhead{}	\\
}
\colnumbers
\startdata
OF024	&	MHO3568	&	20 31 11.98 	&	 +43 5 07.5	&	0.00891	&	0.00538	&	41.4	&	250.0	&	119	&	19	&	Y	&	B	&	[UGPS] 438819307062	&	85	&	16.055	&	14.242	&	12.596	&	11.956	&	16.074	&	13.845	&	12.217	&	10.599	&	8.938	&	7.794	&	3.177	&	G	2	W	A \\
OF025	&	MHO3566	&	20 31 11.87 	&	 +43 5 19.2	&	0.00483	&		&	37.7	&		&	45	&		&	Y	&	S	&	[UGPS] 438819318129	&	0	&		&	19.217	&	16.049	&	16.191	&		&		&		&	15.380	&	11.697	&	10.368	&	4.244	&	G		W	\\
OF026	&	MHO3564	&	20 31 11.60 	&	 +43 5 48.0	&	0.01168	&	0.00864	&	145.8	&	323.0	&	1002	&	421	&	Y	&	B	&	[WISE] J203111.60+430548.0	&	85	&		&		&		&		&		&		&		&	16.677	&	12.154	&	10.820	&	4.227	&			W	\\
\enddata
\tablecomments{(a) Table \ref{tab:MHO_table} is published in its entirety in the 
electronic edition of the {\it Astrophysical Journal}. A portion is 
shown here for guidance regarding its form and content. }
\tablecomments{(b) Column (11): whether the outflow seems to originate from a cluster (Y) or not (N).}
\tablecomments{(c) Column (12): whether the outflow is bipolar (B), single-sided (S) or a single knot/collection of knots (K).}
\tablecomments{(d) Column (26): the point-source catalogues in which the driving source is detected (G for UGPS, 2 for 2MASS, W for AllWISE, A for AKARI, B for Bolocam GPS).}
\end{splitdeluxetable}

\clearpage 
\newpage

\section{Images of the outflows}
\label{sec:appx_images}

\begin{figure}
\figurenum{11}
\plotone{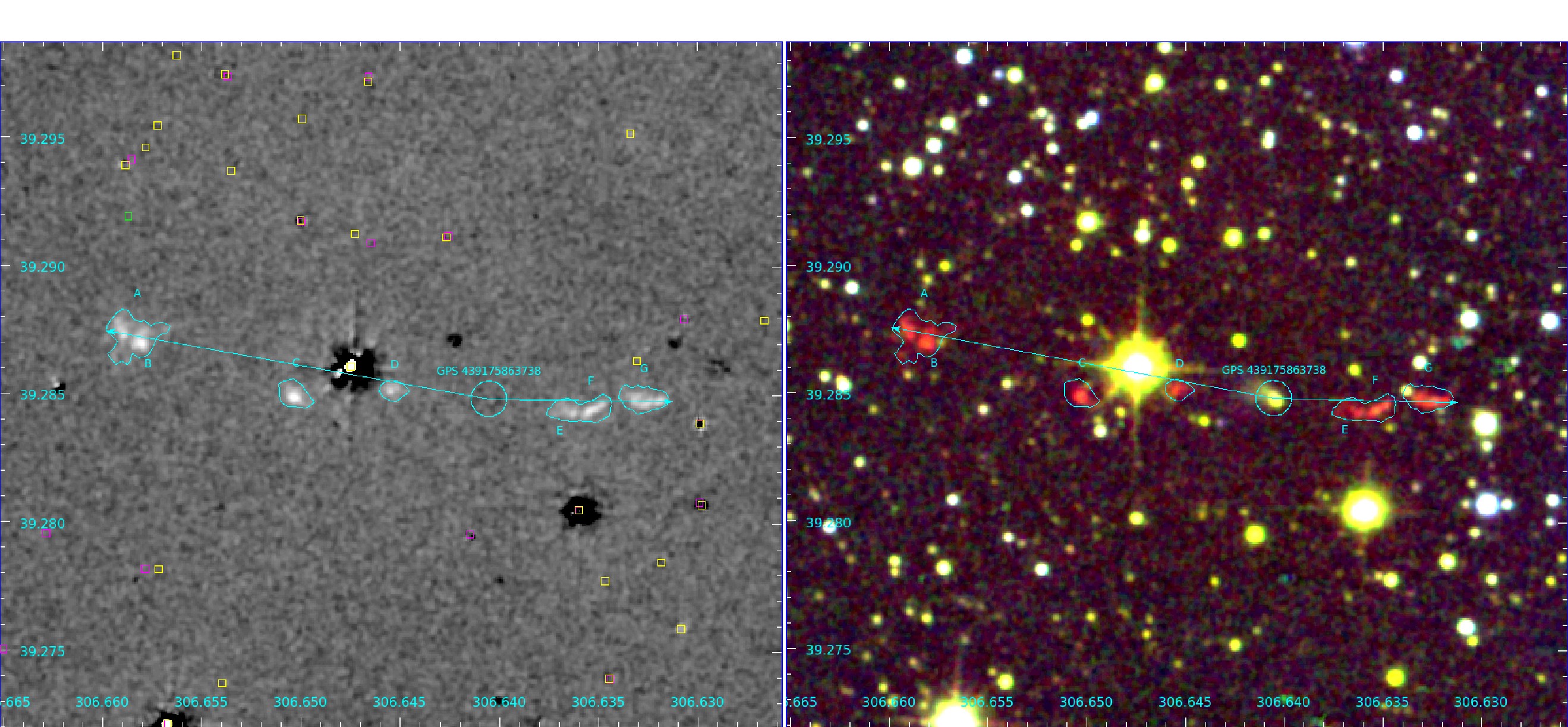}
\caption{OF\,263 (MHO\,3501): A bipolar outflow bisected by a star which has K-band reflection extended along the outflow axis. \label{fig:appx_outflow}}
\end{figure}

In Fig. \ref{fig:appx_outflow} we show an example (OF\,263, MHO\,3501) of an entry from the figure set representing each of our outflows, with the complete set (572 images) being available in the online journal. We show both the \htwo-K (left) and JK\htwo\ (right) images in every case, and the caption contains accompanying details which includes but is not limited to; any relevant Herbig-Haro designation; whether the candidate driving source is a known YSO, and the YSO ID number; and whether the outflow is associated with or nearby to a known cluster, star-forming region (SFR) or \hii\ region. If the outflow is already known as an MHO \citep{davis2007}, we also list all the known MHOs which are included in the outflow and explain the correspondence between our knots and the known MHO ID numbers. 

\clearpage 
\newpage 

\section{Newly discovered clusters}
\label{sec:appx_clusters}

\begin{figure}
\figurenum{12}
\plotone{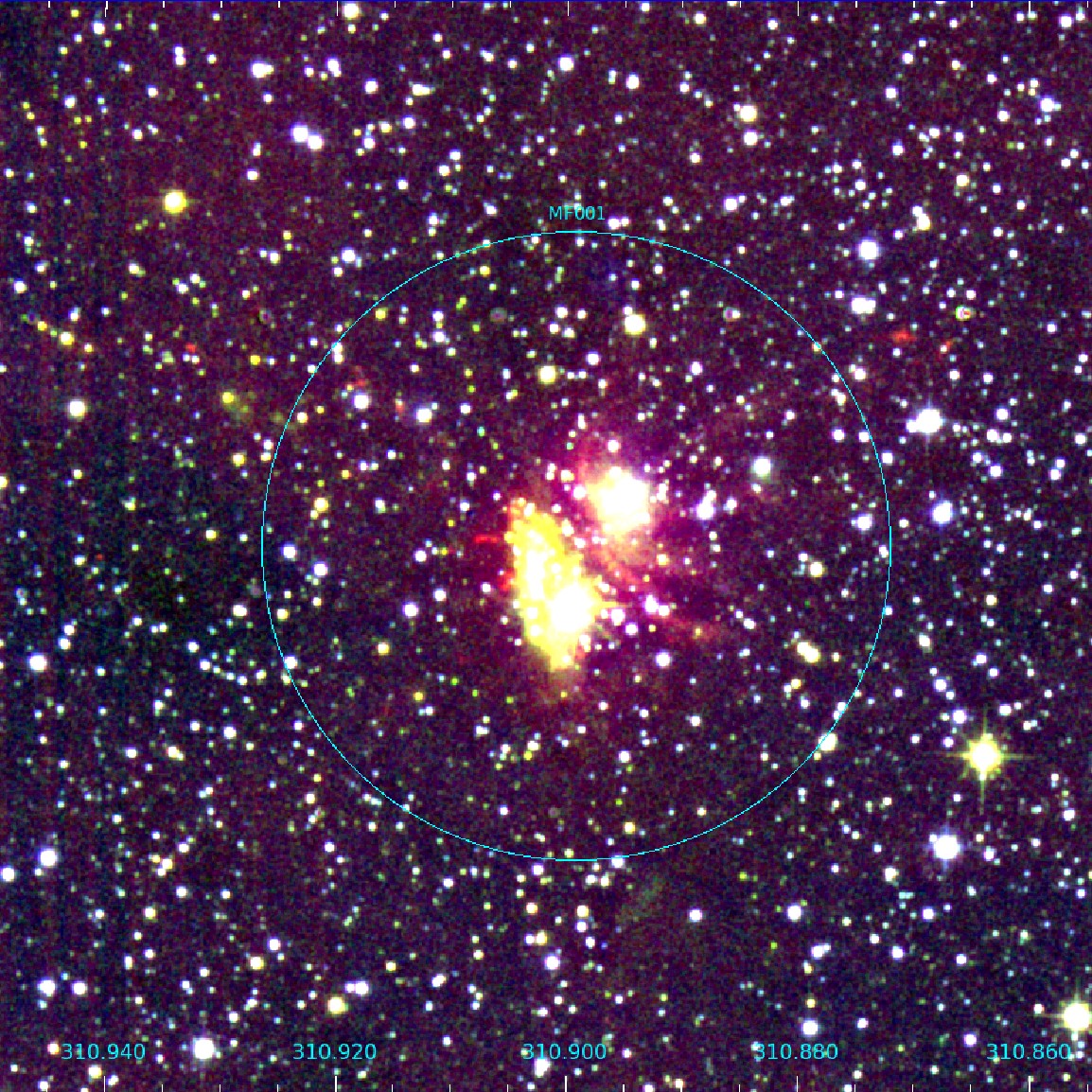}
\caption{MF\,001 Cluster: with $\sim$\,50 NIR members, at 310.89904 $+$44.86583 (1.152\,\,\arcmin radius). A known \hii\ region ([MSX6C]\,G084.1940$+$01.4388), not known as a cluster. Associated with OF\,30.\label{fig:appx_cluster}}
\end{figure}

We report 30 previously-unknown clusters and stellar groups within the Cygnus-X region as a result of this survey. We show images of each of these in a figure set (the complete set of 30 figures is available in the online journal), and Fig. \ref{fig:appx_cluster} is provided as an example of the form and content. In the accompanying caption for each newly-discovered cluster or group, we report additional details such as the position (in J2000), apparent radius (in arcminutes) and number of NIR visible members, as well as a brief description and a listing of any nearby or associated \htwo\ outflows. Note that the ID numbers which appear to be missing are associated with known clusters and hence are excluded from this list. 

\bibliographystyle{aasjournal}
\bibliography{biblio}

\end{document}